\documentclass[aps,prd,twocolumn,nofootinbib]{revtex4-2}

\usepackage{graphicx}
\usepackage{bm}
\usepackage{amsmath}
\usepackage{amssymb}
\usepackage[colorlinks,pdfstartview=FitH]{hyperref}
\hypersetup{linkcolor=blue,citecolor=blue,filecolor=black,urlcolor=blue}

\newcommand\pd{\partial}

\newcommand{\be}{\begin{equation}}
\newcommand{\ee}{\end{equation}}
\newcommand{\bear}{\begin{eqnarray}}
\newcommand{\eear}{\end{eqnarray}}
\newcommand{\ba}{\begin{array}}
\newcommand{\ea}{\end{array}}
\newcommand{\energy}{{\varepsilon}}

\newcommand{\h}{\widehat}
\newcommand{\D}{\mathrm{d}}
\newcommand{\E}{\mathrm{e}}

\newcommand{\I}{{\rm i}}

\newcommand{\mean}[1]{\langle#1\rangle}
\newcommand{\meanlr}[1]{\left\langle#1\right\rangle}
\newcommand{\tr}{{\text{tr}}}
\renewcommand{\vec}[1]{\ensuremath{\mathchoice				
                     {\mbox{\boldmath$\displaystyle\mathbf{#1}$}}
                     {\mbox{\boldmath$\textstyle\mathbf{#1}$}}
                     {\mbox{\boldmath$\scriptstyle\mathbf{#1}$}}
                     {\mbox{\boldmath$\scriptscriptstyle\mathbf{#1}$}}}}
\newcommand{\group}[1]{\relax\ifmmode\mathsf{#1}\else\textsf{#1}\fi}

\def \dphi {\Delta \phi} 
\def \psirp {\Psi_{\rm RP}} 

\usepackage{lineno}

\begin{document}

\author{Matteo Buzzegoli}
\email{mbuzz@iastate.edu}
\affiliation{Department of Physics and Astronomy, Iowa State University, Ames, Iowa 50011, USA
}
\author{Dmitri E. Kharzeev}
\email{dmitri.kharzeev@stonybrook.edu}
\affiliation{Center for Nuclear Theory, Department of Physics and Astronomy,
Stony Brook University, Stony Brook, New York 11794-3800, USA}
\affiliation{Department of Physics, Brookhaven National Laboratory, Upton, New York 11973-5000
}
\author{Yu-Chen Liu}
\email{yu-chen.liu@stonybrook.edu}
\affiliation{Center for Nuclear Theory, Department of Physics and Astronomy,
Stony Brook University, Stony Brook, New York 11794-3800, USA}
\affiliation{Physics Department and Center for Field Theory and Particle Physics, Fudan University, Shanghai 200433, China}
\author{Shuzhe Shi}
\email{shuzhe.shi@stonybrook.edu}
\affiliation{Center for Nuclear Theory, Department of Physics and Astronomy,
Stony Brook University, Stony Brook, New York 11794-3800, USA}
\author{Sergei A. Voloshin}
\email{sergei.voloshin@wayne.edu}
\affiliation{Department of Physics and Astronomy, Wayne State University, 666 W. Hancock, Detroit, Michigan 48201, USA}
\author{Ho-Ung Yee}
\email{hyee@uic.edu}
\affiliation{Physics Department, University of Illinois at Chicago, Chicago, 
Illinois 60607, USA}

\title{ Shear-induced anomalous transport\texorpdfstring{\\}{ }  and charge asymmetry of triangular flow in heavy-ion collisions}

\begin{abstract}
Chiral anomaly implies the existence of non-dissipative transport phenomena, such as the chiral magnetic effect. At second order in the derivative expansion, novel quantum transport phenomena emerge. In this paper, we focus on the anomalous transport driven by a combination of shear, vorticity and magnetic field. We find that the corresponding transport phenomena -- shear-induced chiral magnetic and chiral vortical effects (siCME and siCVE) -- induce characteristic charge correlations among the hadrons produced in heavy ion collisions. We propose the charge asymmetry of triangular flow as a signature of the anomalous transport, and estimate the strength of the signal, as well as the background, using hydrodynamical model simulations. We find that the signal-to-background ratio for the proposed observable is favorable for experimental detection.
\end{abstract}

\maketitle 

{\it{Introduction -}}
The chiral anomaly links the short distance behavior of chiral fermions in quantum field theory to the macroscopic properties of the gauge fields that can possess non-trivial topology. As a result, new kinds of transport phenomena emerge in systems possessing chiral fermions in the presence of magnetic field or vorticity, see~\cite{Kharzeev:2013ffa,Kharzeev:2012ph,Kharzeev:2015znc,Miransky:2015ava,Landsteiner:2016led,Kharzeev:2020jxw} for reviews. The most studied phenomena of this type are the chiral magnetic effect (CME)~\cite{Kharzeev:2004ey,Kharzeev:2007jp,Fukushima:2008xe} and the chiral vortical effect (CVE)~\cite{Kharzeev:2007tn,Erdmenger:2008rm} that describe non-dissipative transport of electric charge along the axis of magnetic field or vorticity in the presence of  chirality imbalance. In addition, at finite vector charge density (e.g. at a finite baryon number density), quantum anomalies induce the axial current in response to both magnetic field and vorticity~\cite{Vilenkin:1980fu,Metlitski:2005pr,Newman:2005as}. The vector and axial currents are coupled by the chiral anomaly, which leads to the emergence of a novel collective excitation, the chiral magnetic wave~\cite{Kharzeev:2010gd}. 

The generation of axial current may be related~\cite{Aristova:2016wxe,Teryaev:2017wlm,Ambrus:2020oiw} to the recently observed polarization of $\Lambda$ hyperons in heavy ion collisions at RHIC~\cite{STAR:2017ckg}. In particular, the measurement of the second order harmonic in the azimuthal angle dependence of longitudinal (along the beam axis) $\Lambda$ polarization~\cite{{STAR:2019erd,Niida:2018hfw}} points towards a substantial role of the shear-induced mechanism of polarization~\cite{Fu:2021pok,Becattini:2021iol}. This raises a question of whether the chiral anomaly may induce a higher harmonic in the azimuthal distribution of electric charge.

\vskip0.3cm
Indeed, such effects were predicted to arise at the second order in the gradient expansion in hydrodynamics as a consequence of chiral anomaly~\cite{Kharzeev:2011ds}. Specifically, the electric current was predicted to possess contributions from shear in the presence of vorticity and magnetic field, and a contribution from the combination of vorticity and magnetic field. The corresponding transport coefficients are proportional to the chiral chemical potential, just like for the CME and CVE, so these effects can be considered as the second-order analogs of CME and CVE. These effects were studied using the effective field theory methods in~\cite{Ammon:2020rvg}.

\vskip0.3cm
{\it{Second order anomalous transport coefficients-}}
The CME and CVE are of first order in the hydrodynamic gradient expansion, and the corresponding transport coefficients can be derived in the framework of hydrodynamics by imposing the non-negativity of entropy production~\cite{Son:2009tf}. At second order, there appear additional transport coefficients that have been classified in Ref.~\cite{Kharzeev:2011ds}. The relations between these transport coefficients have been derived from the {\it absence} of entropy production that stems from the time reversal invariance~\cite{Kharzeev:2011ds}. 
\vskip0.3cm

In this paper we will focus on the contributions to electric current that arise from the combination of shear and vorticity or magnetic field~\cite{Kharzeev:2011ds}:
\be\label{cme2}
j^\mu_{(2)} = \xi_1 \sigma^{\mu\nu}\omega_\nu + \xi_2 Q \sigma^{\mu\nu}B_\nu,
\ee
where $\sigma^{\mu\nu}= {1\over2} (\partial^\mu_\perp u^\nu + \partial^\nu_\perp u^\mu)$ is the transverse shear tensor ($u^\mu$ is the fluid velocity, and $\partial_\perp^\mu$ is the gradient perpendicular to $u^\mu$), $\omega^{\mu}={1\over 2}\epsilon^{\mu\nu\alpha\beta}u_\nu\partial_\alpha u_\beta$ is vorticity, and $B^\mu$ is magnetic field. 
We will refer to the first term in (\ref{cme2}) as the shear-induced Chiral Vortical Effect (siCVE), and to the second term as the shear-induced Chiral Magnetic Effect (siCME).\footnote{This effect, introduced in ~\cite{Kharzeev:2011ds}, is referred to as shear-induced Hall effect in Ref.~\cite{Ammon:2020rvg}.}
\vspace{-0.3cm}
\begin{figure}[!hptb]
\hspace*{-2cm}
   \includegraphics[width=0.7\textwidth]{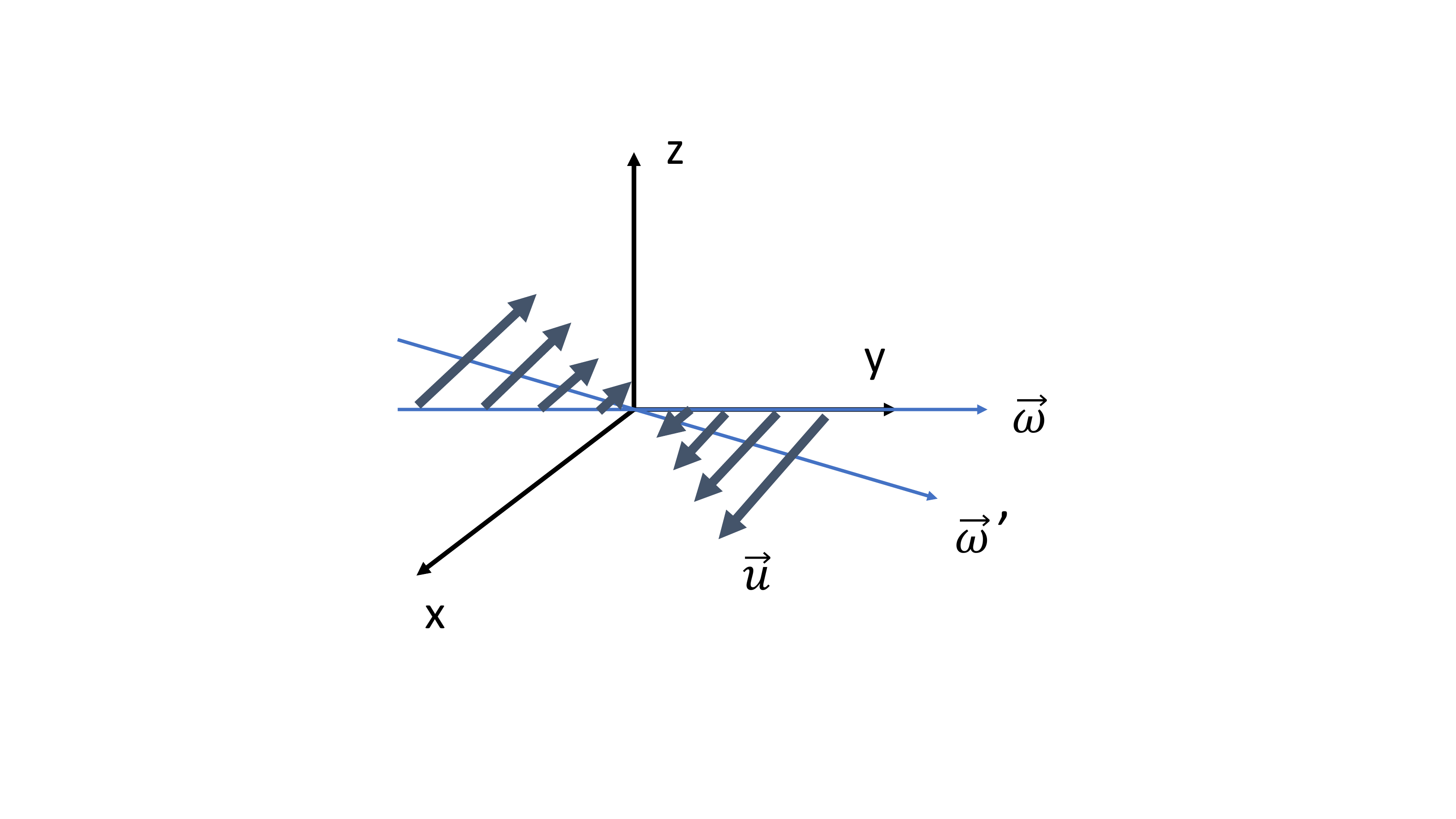}
  \vspace*{-0.7cm}
    \caption{The tilting of vorticity in a shear flow. In the presence of a shear flow $\sigma^{xy}$ (the fluid velocity ${\vec u}$ is along ${\vec x}$ with ${\partial u^x / \partial y} \neq 0$), the vortex immersed in the flow and originally pointing in the direction ${\vec\omega} \sim {\vec y}$ gets tilted and points in the direction ${\vec \omega}^\prime$, thus acquiring a component along ${\vec x}$.}
    \label{fig:vortex}
\end{figure}

\vskip0.3cm

What is the microscopic origin of the phenomena encoded in (\ref{cme2})? It is well known that the anomaly relation, and thus the expression for the CME current
\be\label{cme}
{\bf j} = \frac{e^2}{2\pi^2} \mu_5 {\bf B}
\ee
are exact \textit{at the operator level}.  Nevertheless, when the expectation value of this operator relation is taken over a physical state, there may well appear corrections arising from the renormalization of operator quantities that enter (\ref{cme}), see e.g.~\cite{Anselm:1989gi} and discussion in~\cite{Adler:2004qt}. In particular, the magnetic field in the medium can be renormalized by interactions. Moreover, if the shear (and/or vorticity) are present in the medium, they can rotate the orientation of an effective magnetic field by generating a component of the field in the direction perpendicular to initial ${\bf B}$.

To illustrate this argument, let us consider a vortex immersed in the flow and aligned initially along the axis $y$, with ${\vec \omega} \sim {\hat{\vec y}}$. The shear flow with $\sigma^{xy} \sim \omega_z$ will rotate the axis of the vortex in the $(x, y)$ plane, creating a component of an effective vorticity along the axis $x$, see Fig.~\ref{fig:vortex}. This ``tilting" of vorticity in shear flows has been extensively studied in hydrodynamics, see~\cite{kawahara1997wrap} and references therein. Perhaps the most spectacular manifestation of vorticity tilting in Nature is the emergence of tornadoes in ``supercell" thunderstorms, see~\cite{dahl2017tilting} for a review.

The ``conventional" first order chiral vortical effect will then create the current along the $x$ axis. Therefore, the second order anomalous transport phenomenon can be understood in terms of the modification of vorticity (or magnetic field) by the back-reaction of the medium. 

\vskip0.3cm
The values of the second-order transport coefficients $\xi_i$ had been evaluated at strong and weak coupling through holography and chiral kinetic theory, respectively. These computations will be briefly summarized below. One can also write down the general Kubo relations for these coefficients that will be described below as well.

\vskip0.3cm

{\it Transport coefficients at strong coupling -} The value of $\xi_1$ has been computed by holographic methods in 
$N=4$ supersymmetric Yang--Mills (SYM) theory~\cite{Erdmenger:2008rm}. Although the conformal $N=4$ SYM and QCD are clearly not the same theories, we may estimate $\xi_1$ for QCD basing on the $N=4$ SYM result.  We thus get 
\be
\xi_1 =  -{N_F N_c\over \sqrt{3}\pi^3}{\mu_A\mu\over T}\quad ({\rm strong\,coupling})
\ee
where $\mu$ and $\mu_A$ are the chemical potentials of the vector and axial charges, respectively. For numerical estimates we will assume that the number of light quark flavors $N_F=3$. 

Considering parity and charge conjugation symmetries, $\xi_2$ is proportional to $\mu_A$ only, and we estimate  
\be
\xi_2 = -{N_FN_c\over \sqrt{3}\pi^3}{\mu_A\over T}\quad ({\rm strong\,coupling})
\ee

It is interesting to note that these transport coefficients are the result of interplay between chiral anomaly and the dissipative dynamics of the plasma represented by shear viscosity and conductivity.
This physics seems to be unique for these transport terms, among other possible second order terms. This feature will be important in our computation of these transport coefficients in weakly coupled regime, and also in the derivation of Kubo relations based on the Zubarev  approach.
\vskip0.3cm

{\it Weakly coupled regime: the chiral kinetic theory -}
To demonstrate the universal nature of the discussed phenomenon, let us now discuss how it emerges at weak coupling. For this purpose we will present a derivation of shear-induced anomalous transport using the Chiral Kinetic Theory (CKT)~\cite{Son:2012wh,Stephanov:2012ki,Chen:2012ca}. 

Using the covariant fermion Wigner function
\begin{equation}
  W_{ab}(x,p) = \int d^4y\,e^{\frac{-i}{\hbar}p\cdot y}\,
  \langle \bar{\psi}_b(x)e^{y\cdot\overleftarrow{\nabla}}
  e^{-y\cdot \nabla}\psi_a(x) \rangle\,,
\end{equation}
where $\nabla_\mu \psi = \left( \partial_\mu + i \,Q\, A_\mu/\hbar \right) \psi$, we can express the vector current in the form $j^\mu = \int \frac{d^4 p}{(2\pi)^4} {\rm Tr} \left[ \gamma^\mu W(x,p) \right]$. 

Dirac equation for charged massless fermions in a constant  electromagnetic field leads to the following equation for the Wigner function:
\begin{equation}
    \gamma^\mu \left( p_\mu + \frac{i\hbar}{2} \Delta_\mu \right) W(x,p)=0, 
\end{equation}
where $\Delta_\mu = \partial_\mu - Q\,F_{\mu\lambda}\partial_p^\lambda$; we regard the electromagnetic field $F^{\mu\nu}$ as the first order quantity in the derivative expansion.

\vskip0.3cm
We are interested in the vector current that represents the sum of right-handed and left-handed chiral currents: $j^\mu = j_+^\mu + j_-^\mu $. The right- and left-handed currents of charge $Q$ fermions with (dual) electromagnetic field $\tilde F^{\mu\nu}$ are given in CKT by
\begin{eqnarray}
j^\pm_\mu
&=& \int \frac{d^4 p}{(2\pi)^4} \left[ 4\pi \delta(p^2) p_\mu f^\pm 
+\tilde{\mathcal{J}}^\pm_\mu 
+\tilde{\tilde{\mathcal{J}}}^\pm_\mu \right]\,,
\label{eq:ckt_current}
\end{eqnarray}
with
\begin{eqnarray}
\tilde{\mathcal{J}}^\pm_\mu &\equiv& 4\pi \hbar  \delta(p^2) \left\{ \mp \frac{Q}{p^2}\widetilde{F}_{\mu\sigma}p^\sigma f^\pm 
\pm \Sigma^n_{\mu\rho}\Delta^\rho f^\pm\right\} \,,
\nonumber\\
\tilde{\tilde{\mathcal{J}}}^\pm_\mu
&\equiv&
 \mp p_\mu \frac{1}{p^2} \frac{\hbar}{2 p\cdot n}\epsilon_{\alpha\beta\rho\sigma} p^\alpha n^\beta \Delta^\rho\tilde{\mathcal{J}}_\pm^\sigma \nonumber\\
&&
\pm \frac{\hbar}{2 p\cdot n}\epsilon_{\mu\nu\rho\sigma}n^\nu \Delta^\rho \tilde{\mathcal{J}}_\pm^\sigma \,,
\end{eqnarray}
where $f^\pm$ is the distribution function for right(left)-handed particles, $\Sigma_n^{\mu\nu}\equiv \epsilon^{\mu\nu\rho\sigma}p_\rho n_\sigma/(2p\cdot n)$ is a spin tensor for chiral fermions and $n^\mu$ is a unit time-like frame that satisfies $n^2 = 1$. The second-order expressions for the current have also been derived, both for the case of background electromagnetic field~\cite{Yang:2020mtz} and for a  curved space-time in Ref.~\cite{Hayata:2020sqz}. 
\vskip0.3cm

According to the  analysis~\cite{Kharzeev:2011ds}, the shear-induced second order terms are given by (\ref{cme2}). We find that such terms do not arise from $\tilde{\tilde{\mathcal{J}}}^\pm_\mu$. 
This is consistent with the qualitative analysis given above that indicates that these shear-induced terms originate from the medium modifications of the distribution function. We will see below that once these modifications are taken into account in the first order distribution function $f^\pm_{(1)}$, the shear-induced current indeed emerges from 
 $\tilde{\mathcal{J}}^\pm_\mu$.

We assume, as usual, that the distribution function depends on the linear combination of quantities that are collisionally conserved at local equilibrium, so that the detailed balance condition can be satisfied.
For the case of Fermi--Dirac distribution we thus obtain~\cite{Liu:2020flb}
\begin{eqnarray}
\label{eq:EqDistFunc}
f_\mathrm{eq}^\pm = \left[ e^{(p\cdot \beta \mp \frac{\hbar}{2} \Sigma_n^{\alpha\beta}\gamma_{\alpha\beta} - \alpha^\pm)} +1 \right]^{-1}\,,
\end{eqnarray}
where $\alpha^\pm=\mu^\pm/T$ and $\beta_\mu = u_\mu/T$ are respectively the temperature-scaled chemical potential and fluid velocity, and $\gamma_{\alpha\beta} =- \frac{1}{2}(\partial_{\alpha}\beta_{\beta}-\partial_{\beta}\beta_{\alpha})$ is thermal vorticity.
The above distribution function agrees with the one obtained within the exact density matrix approach~\cite{Palermo:2021hlf} at first order in vorticity;
we can thus use it to describe the effects at  first order in vorticity. An important point for us is that there is no shear-induced term in $f^\pm_{\rm eq}$, which means that we cannot derive second order shear-induced terms in $\tilde{\mathcal{J}}^\pm_\mu$ using the distribution function (\ref{eq:EqDistFunc}). 
\vskip0.3cm

In order to include the shear contributions, we need to consider the viscous corrections to the distribution function. For this purpose, we employ the moment expansion method~\cite{Denicol:2012cn} to formulate the non-equilibrium distribution:
\begin{equation}
\begin{split}
f^\pm =\,& f_\mathrm{eq}^\pm + f_\mathrm{eq}^\pm (1-f_\mathrm{eq}^\pm) 
	\Big( 
    \lambda_{\Pi}^\pm \Pi
		+	\lambda_{\nu}^\pm \nu_{\pm}^\mu p_\mu
		+	\lambda_{\pi}^\pm \pi^{\mu\nu} p_\mu p_\nu 
	\Big), \label{eq.non_equilibrium_distribution}
\end{split}
\end{equation}
and compute the shear-induced chiral transport coefficients.
Here, $\lambda_{X}^{\pm}$ are polynomials of $u\cdot p$, with coefficients being functions of $T$ and $\alpha^\pm$. They are determined by matching the energy-momentum stress tensor from its microscopic integral  representation to the corresponding macroscopic viscous terms. Following this approach,  we find that (the details of the derivation are presented in the Supplementary Material~\ref{app:1}~\cite{SupplementaryMaterial}):
\begin{align}
\begin{split}
\xi_1 \approx\,& 
- 0.62 {\eta \over s}{\mu_A\mu\over T}
= -0.05{\mu_A\mu\over T}\;({\rm weak\,coupling}),\\
\xi_2 \approx\,& 
-6.70{\eta \over s}{\mu_A\over T}
= -0.53{\mu_A\over T}\quad\;\, ({\rm weak\,coupling}),
\end{split}\label{wc-tr}
\end{align}
where $\eta$ is a shear viscosity, and $s$ is an entropy density. 

While the relations (\ref{wc-tr}) have been obtained within the chiral kinetic theory that is applicable at weak coupling, the second equalities in (\ref{wc-tr}) are based on the assumption $\eta/s=1/(4\pi)$ that follows from holography at strong coupling and is favored by the data. The use of weak coupling value of $\eta/s$ would yield substantially bigger values of the transport coefficients $\xi_1$ and $\xi_2$ -- so the values (\ref{wc-tr}) can be considered as lower bounds on these quantities at weak coupling. 
Let us note that the $\xi$ coefficients have been computed in Ref.~\cite{Hidaka:2018ekt} using relaxation time approximation, as well as using moment expansion method in Ref.~\cite{Shi:2020htn} for a single-component fluid.
\vskip0.3cm

{\it Kubo relations in the Zubarev approach -}
It is important to establish the general Kubo formulae for the transport coefficients of siCME and siCVE. For a relativistic quantum system, Kubo relations
can be obtained using the linear response theory in the Zubarev
formalism for the non-equilibrium statistical operator~\cite{Hosoya:1983id,Becattini:2019dxo}.
 
In this formalism, a covariant form of the local thermal equilibrium statistical operator
is obtained by maximizing the total entropy at fixed energy-momentum
density~\cite{Zubarev:1966,Zubarev:1979,vanWeert1982,Zubarev:1989su,Morzov:1998}.
In the presence of vorticity the statistical operator around a point $x$ can be
approximated as~\cite{Becattini:2014yxa}:
\begin{equation*}
\h{\rho}\simeq \frac{1}{Z} \exp\left\{-\beta(x)\cdot \h{P} +\h{B}_\omega + \h{B}_{\rm D} \right\}
\end{equation*}
with $\h{P}$ the total momentum of the system and
\begin{equation*}
\begin{split}
\h{B}_\omega = & - \beta(x)\omega_\rho(x) \h{J}_x^\rho, \quad
\h{B}_{\rm D} = \int_\Omega\D\Omega\, \h{T}^{\mu\nu}\nabla_\mu\beta_\nu,
\end{split}
\end{equation*}
where $\h{J}_x$ is the angular momentum of the system evaluated around the point $x$
and $\Omega$ is the region of space-time enclosed by the two hyper-surfaces at the initial thermalization time 
and at the present time, and by the time-like hyper-surface at their boundaries. The operator
$\h{B}_\omega$ describes the non-dissipative effects related to vorticity, while
$\h{B}_{\rm D}$ describes the dissipative effects. In particular, the latter contains
the contribution from the shear tensor that can be written as
$\h{B}_\eta=\int_\Omega\D\Omega\, \h{T}^{\mu\nu}\beta\sigma_{\mu\nu}$.
\vskip0.3cm

The siCVE is obtained by evaluating the current
$j^\mu(x)=\tr[\h\rho\,\h{j}^\mu(x)]$ as a linear response to $\h{B}_\omega$
and $\h{B}_D$ and considering the term of order $\h{B}_\omega\times\h{B}_\eta$.
Using the linear response theory as in~\cite{Huang:2011dc,Becattini:2015nva,Buzzegoli:2017cqy,Buzzegoli:2018wpy,Harutyunyan:2018cmm,Buzzegoli:2020ycf,Harutyunyan:2021rmb}
and expressing the correlators in terms of the three-point retarded Green function~\cite{Grossi:2014,Evans:1990qh,Evans:1991ky}
\begin{equation*}
\begin{split}
\I &G^{\rm R1}_{\h{O},\h{X},\h{Y}}(x;\,x_1,\,x_2) = \\
    &\theta(t-t_1) \theta( t_1 -t_2)\meanlr{\left[\left[\h{O}(x),\, \h{X}(x_1)\right], \h{Y}(x_2)\right]}_T\\
	&+\theta(t-t_2) \theta( t_2 -t_1)\meanlr{\left[\left[\h{O}(x),\,\h{Y}(x_2)\right],  \h{X}(x_1)\right]}_T,
\end{split}
\end{equation*}
we obtain (see Supplementary Material~\ref{app:2}~\cite{SupplementaryMaterial} for detailed derivation)
\begin{equation*}
\begin{split}
\Delta_{\omega\eta} j^\mu(x) =& \frac{2 \omega_\nu(x)}{\beta(x)}
    \int \D^4 x_1 \int \D^4 x_2 \int_{-\infty}^{t_2}\!\!\!\D\theta_2\, \beta(x_2)\sigma^{\mu\nu}(x_2)\\
	&\times  (x_1-x)^y\, \I\, G^{\rm R1}_{\h{j}^y,\h{T}^{tz},\h{T}^{xy}}\left(x;\, x_1,\, (\theta_2, \vec{x}_2)\right),
\end{split}
\end{equation*}
where we denoted by $\mean{\h{O}}_T$ the trace with the homogeneous statistical operator in the
local rest frame with temperature $T=1/\beta(x)$. Notice that since the effects that we seek to describe require
breaking of parity, this statistical operator must also contain a chiral imbalance.

We can move the shear tensor out of the integration by studying the perturbations
with respect to the equilibrium. For a fluid in the hydrodynamic regime, only the
perturbations with small frequency and small wave vector contribute to the integral.
In that case, following~\cite{Becattini:2019dxo}, we obtain $\Delta_{\omega\eta} j^\mu(x)
=\sigma^{\mu\nu}(x) \omega_\nu(x) \xi_1$ with
\begin{equation}
\label{eq:Xi1KuboFormula}
\xi_1 = \lim_{p,q\to 0} 2 \frac{\pd}{\pd q^0}\frac{\pd}{\pd p^y}
	\text{Im} G^{\rm R1}_{\h{j}^y,\h{T}^{tz},\h{T}^{xy}}(p,q)
\end{equation}
where $G^{\rm R1}(p,q)\!=\!\int\! \D^4 x_1 \D^4 x_2\, \E^{-\I(p\cdot x_1 + q\cdot x_2)} G^{\rm R1}(0;x_1,x_2) $.
Moreover, comparing the known Kubo formulas of the CME and CVE, we see that
we can obtain $\xi_2$ from $\xi_1$ replacing $\h{T}^{tz}$ with
$(\beta/2)\h{j}^z$, that is
\begin{equation}
\label{eq:Xi2KuboFormula}
\xi_2 = \lim_{p,q\to 0} \beta \frac{\pd}{\pd q^0}\frac{\pd}{\pd p^y}
	\text{Im} G^{\rm R1}_{\h{j}^y,\h{j}^z,\h{T}^{xy}}(p,q) .
\end{equation}

\vskip0.3cm

\begin{figure}[!hptb]\centering
    \includegraphics[width=0.4\textwidth]{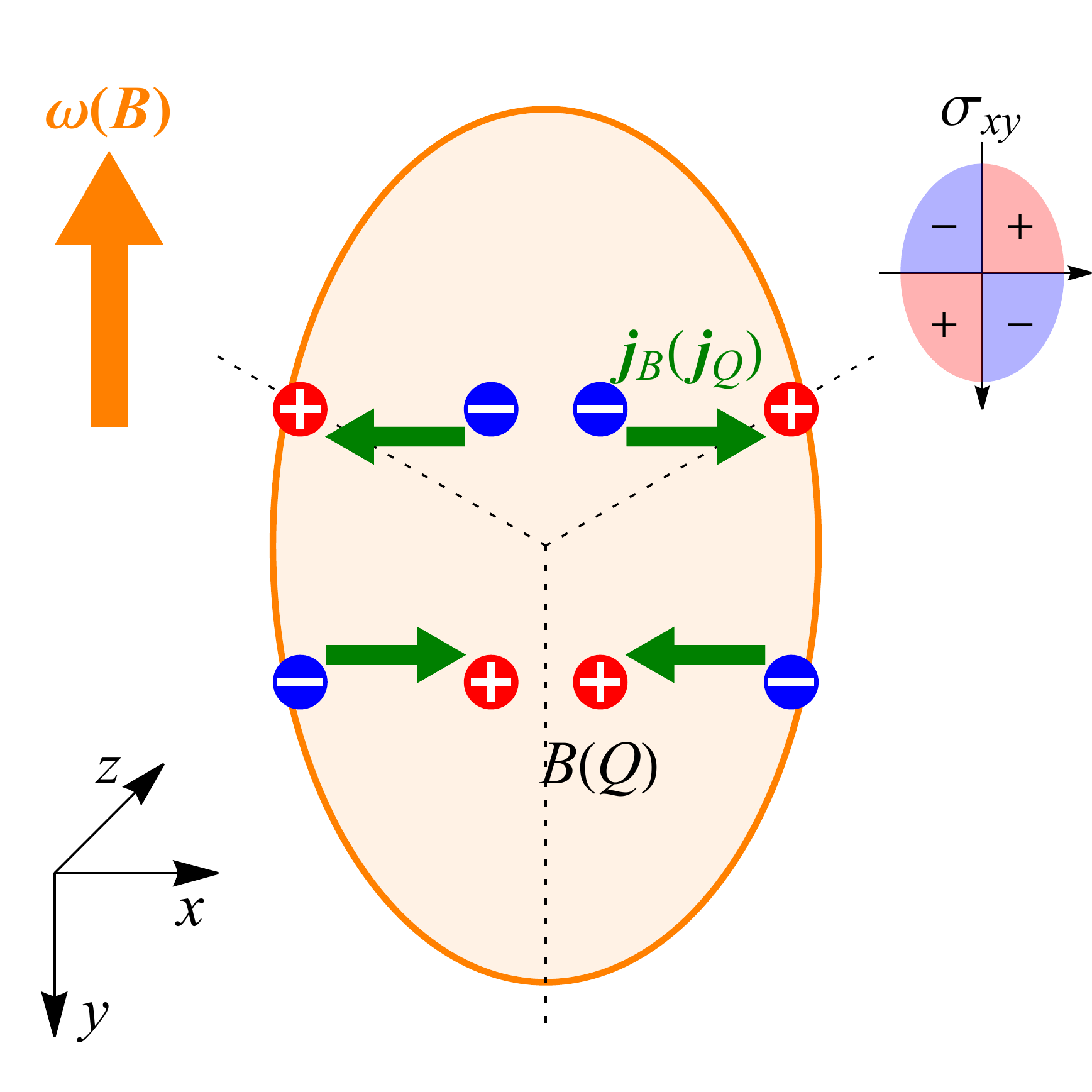}
    \caption{Illustration of the shear-induced chiral vortical and magnetic effects; see text for the description.}
    \label{fig:illustrate}
\end{figure}

{\it Experimental observable: charge dependent fluctuations of \texorpdfstring{$a_3$}{a3} - }
Let us now discuss the experimental signatures of siCME and siCVE. Let us assume that the beam direction of the colliding ions is along the axis ${\bf z}$, and the axis ${\bf x}$ lies in the reaction plane, see Fig.~\ref{fig:illustrate}. The elliptical flow of the expanding quark-gluon plasma then induces the dependence of the fluid velocity component $u_x$ on $y$, and thus the shear $\sigma_{xy}$ of the sign that is indicated in the insert of Fig.~\ref{fig:illustrate}. The axes of vorticity and magnetic field are aligned perpendicular to the reaction plane, anti-parallel to ${\bf y}$. The resulting siCME and siCVE currents are thus directed parallel or anti-parallel to ${\bf x}$, depending on the sign of $\sigma_{xy}$, as shown in Fig.~\ref{fig:illustrate}.

\begin{figure}[!hptb]\centering
    \includegraphics[width=0.4\textwidth]{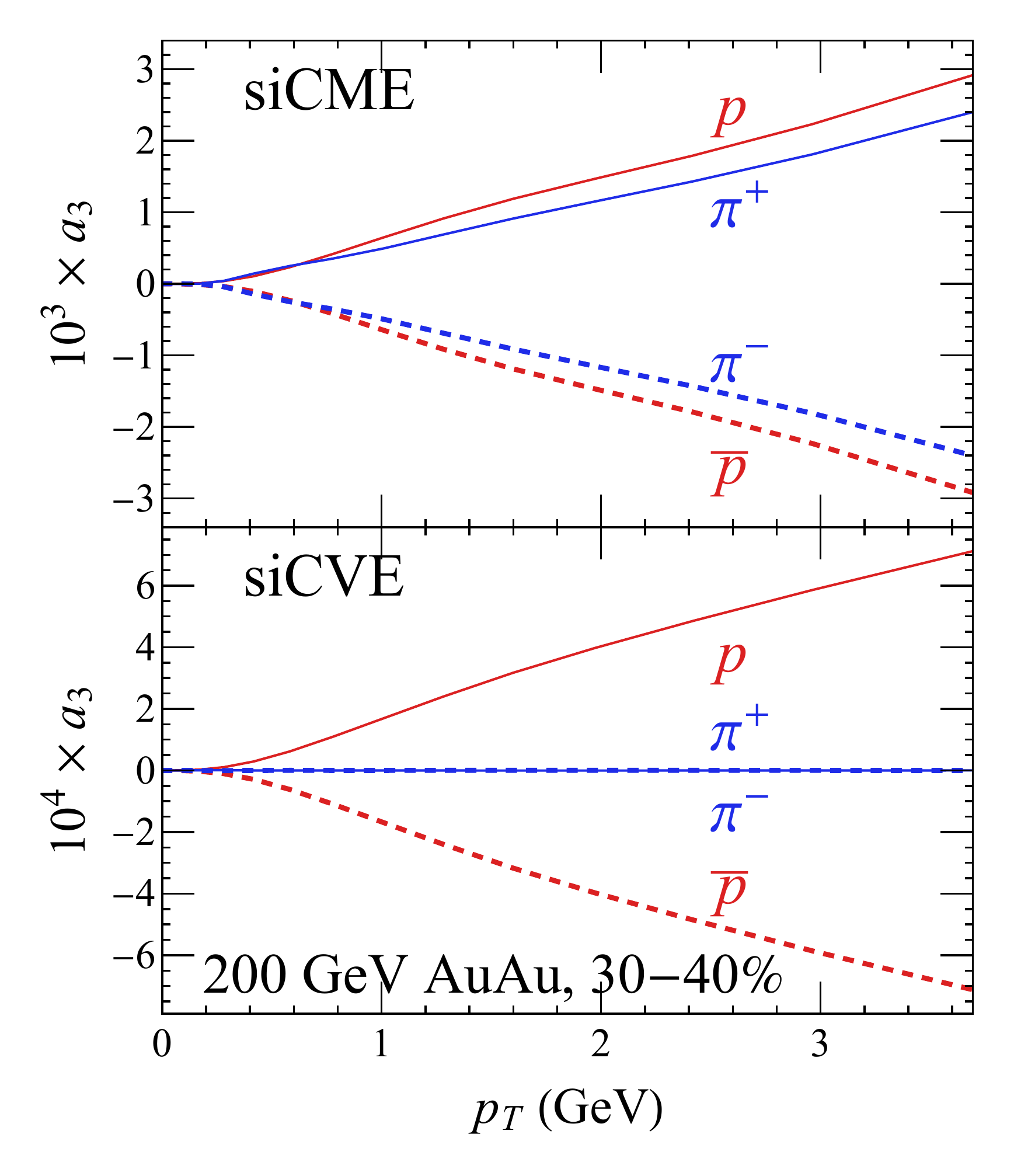}
    \caption{(Upper) Transverse momentum dependence of third-harmonic charge separation for $\pi^+$, $\pi^-$, $p$, and $\bar p$ due to the shear-induced chiral magnetic effect. (Lower) Same as the upper panel but for the shear-induced chiral vortical effect.}
    \label{fig:pt_dependence}
\end{figure}

\begin{figure}[!hptb]\centering
    \includegraphics[width=0.4\textwidth]{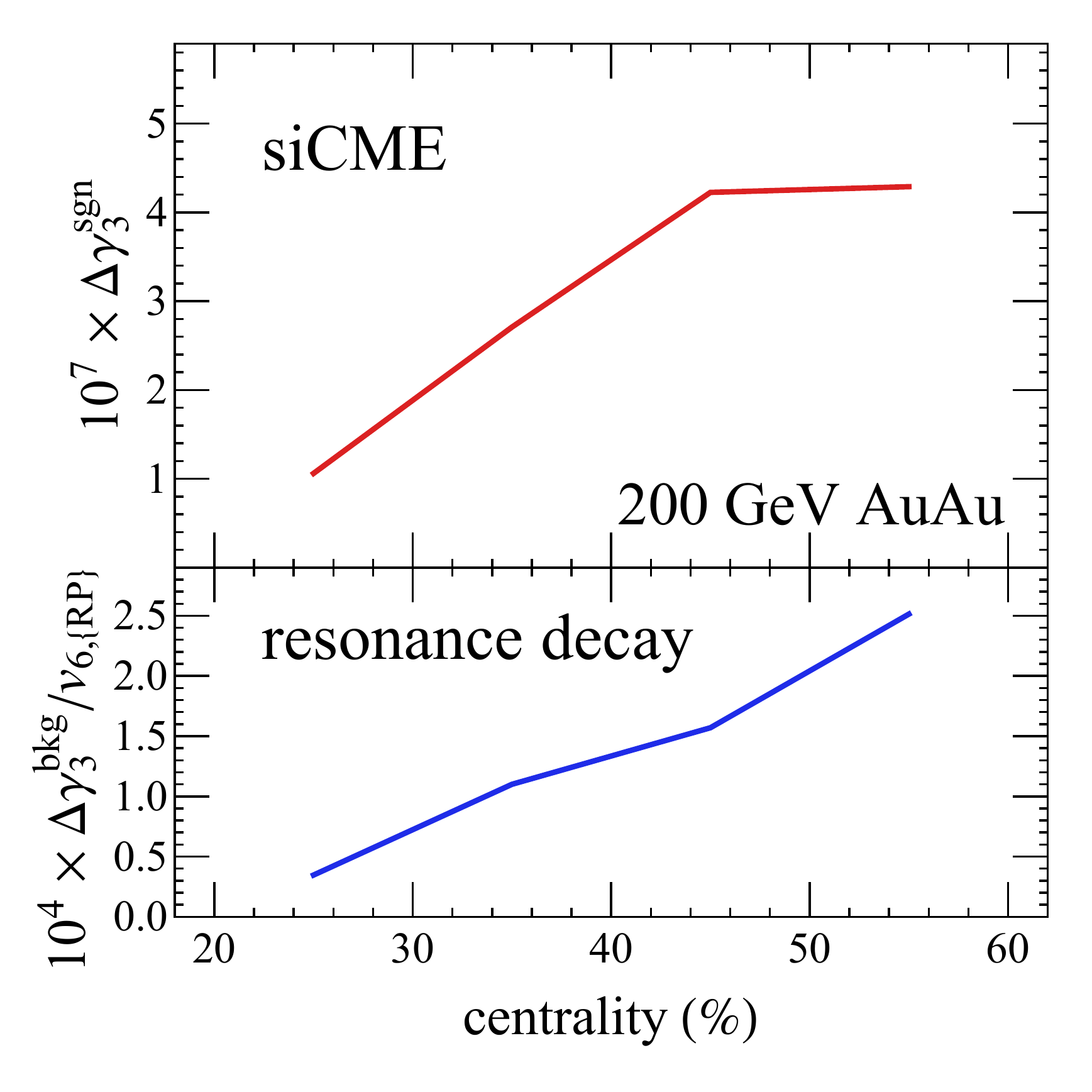}
    \caption{Centrality dependence of two-point correlation $\Delta\gamma_3$ induced by shear-induced chiral magnetic effect (upper) and resonance decay background (lower).}
    \label{fig:centrality_dependence}
\end{figure}

\vskip0.3cm
 As a result, the siCME, i.e., the current term $\xi_2 \sigma^{\mu\nu}B_\nu$, would lead to the triangular distortion of the particle momentum distribution, that will be different for positive and negative particles. The sign of this distortion for positive and negative particles will fluctuate event-by-event, reflecting the fluctuations of the chiral chemical potential $\mu_A$. This is similar to the charge-dependent dipole distortion of the momentum distribution induced by the ``conventional" CME.

The azimuthal particle distribution in heavy ion collision is often
parameterized with Fourier series:
%
\begin{equation}
\frac{dN}{d\phi} \propto 1+\sum_{n=1}^\infty [ 2 v_n \cos(\dphi)
  + 2a_n\sin(\dphi)],
\end{equation}
where $\dphi = \phi-\psirp$ is the emission angle relative to the
reaction plane. Coefficients $a_n$ are zero if parity is conserved in
the collision. Experimental search for the conventional CME is focused
on measuring correlators sensitive to the product $\mean{a_{1,\alpha}
  a_{1,\beta}}$  where $\alpha$ and $\beta$ denote the charge of the
particles~\cite{Voloshin:2004vk}. 
It is usually done by measuring the so-called ``gamma''
correlator
$\gamma_1^{\alpha\beta} = \mean{\cos(\phi_\alpha + \phi_\beta - 2\Psi_\textsc{RP})}$. 
   
The charge-dependent triangular distortions of the momentum distributions can be detected by 
 the third order sine harmonics $a_3 \equiv \langle \sin(3\phi-3\Psi_\textsc{RP}) \rangle$ evaluated for
particles with positive $a_3^+$ and negative $a_3^-$ charges. Namely, $a_3^+ = -a_3^- \neq 0$ in each event, where the sign of $a_3^+$ depends on the sign of chiral imbalance.

In analogy to the ``conventional" CME observables, we thus define the two-particle correlator of third order harmonics, $\gamma_3^{\alpha\beta} \equiv \langle \cos(3\phi_\alpha + 3\phi_\beta - 6\Psi_\textsc{RP})\rangle$. This correlator for same-sign(\textsc{SS}) and opposite-sign(\textsc{OS}) pairs responds to the effect as follows: $\gamma_3^\textsc{SS} = \gamma_{3,\textrm{bkg}}^\textsc{SS} - (a_3^+)^2$ and $\gamma_3^\textsc{OS} = \gamma_{3,\textrm{bkg}}^\textsc{OS} + (a_3^+)^2$, where $\gamma_{3,\textrm{bkg}}$ represents the background contributions. Therefore, the difference $\Delta\gamma_3\equiv \gamma_3^\textsc{OS}-\gamma_3^\textsc{SS}$ should be sensitive to the siCME. 

For the siCVE, induced by the current term $\xi_1 \sigma^{\mu\nu} \omega_\nu$, the analysis is very similar, but one expects it to lead mostly to the separation of baryons and antibaryons, just as for the CVE~\cite{Kharzeev:2010gr}. 
\vskip0.3cm

To estimate the signal, we use the AVFD simulation framework~\cite{Jiang:2016wve,Shi:2017cpu,Shi:2019wzi} to evaluate the vector and axial-vector charge evolution on top of a realistic hydro background with axial charge initial condition $|n_A/s| = 0.1$ (equivalently, $|\mu_A/T| \sim 1$) and magnetic field lifetime $\tau_B = 1$~fm. The magnitude and the spatial distribution of the initial chirality imbalance are set to be the same as in the CME simulation. Even if the size of topological fluctuations is small, the assumption of a uniform distribution may still capture  the average effect resulting from the random diffusion of topological charge that leads to Chern--Simons number of order $\sqrt{N}$, where $N$ is the number of sphalerons. The initial profile of the magnetic field is computed from the initial proton distribution of the colliding nuclei. The bulk evolution starts from the event-averaged Monte Carlo Glauber initial conditions, followed by solving 2+1D second-order viscous hydrodynamic equations with MUSIC~\cite{Schenke:2010rr,Gale:2013da}.

We focus on top energy $\sqrt{s_{NN}} = 200$ GeV Au+Au collisions and compute the observables proposed above for the detection of shear-induced chiral effects.
As an example, we take an event with an excess of  right-handed particles ($n_5>0$), and show the transverse momentum dependence of the $a_3$ moments. The cases of siCME and siCVE are shown in the upper and lower panels of Fig.~\ref{fig:pt_dependence} respectively.
We observe a $\mathcal{O}(10^{-3})$ difference between $a_3^{\pi^+}$ and $a_3^{\pi^-}$ ($a_3^{p}$ and $a_3^{\bar{p}}$) due to the siCME(siCVE) effect, and the separation increases with transverse momentum $p_T$. 
We find that the contribution from ``conventional" CME to $a_3$ is an order of magnitude smaller than the contribution from siCME. 

The amplitude of the signal in $a_3$ is smaller than the CME $a_1$ charge separation by an order of magnitude, as appropriate for a second-order effect in the hydrodynamical derivative expansion. It is thus especially important to estimate the non-chiral effect of the background on $\Delta \gamma_3$ before a conclusion on observability of siCME and siCVE  in heavy-ion collisions can be reached.

 \vskip0.3cm

To estimate the background from resonance decays, we sample the resonances according to their transverse distribution, $\frac{\mathrm{d}N^\mathrm{res}} {\mathrm{d}p_T\mathrm{d}\phi}$, and collect the decay particle pairs that fall into the kinematic region of interest that is chosen to be $0.5<p_T<2$~GeV and $|\eta|<0.5$. The lower $p_T$ cut is chosen to enhance the signal and suppress the background. We include the two-particle decay of $K^0_S$, $\rho^0$, and $\omega$ and three-particle decay of $K^0_L$, $\eta$, and $\omega$ particles. 

Noting that for resonance decays $\langle \cos(3\phi_\alpha + 3\phi_\beta - 6\Psi_\textsc{RP})\rangle \approx \langle \cos(3\phi_\alpha + 3\phi_\beta - 6\phi^\mathrm{res}) \cos(6\phi^\mathrm{res}-6\Psi_\textsc{RP})\rangle
\approx \langle \cos(3\phi_\alpha + 3\phi_\beta - 6\phi^\mathrm{res})\rangle \, v_{6,\textsc{RP}}^\mathrm{res}$ and similarly for local-charge conservation~\cite{Schlichting:2010qia,Liao:2010nv}, we expect that the background contribution in $\Delta \gamma_3$ is proportional to the sixth-order flow harmonic with respect to the reaction plane (or, the second-order event plane). In Fig.~\ref{fig:centrality_dependence}, we present the centrality dependence of siCME signal ($\Delta\gamma_3^\mathrm{sgn}\equiv2a_3^2$) and the background induced by resonance decay, with the latter scaled by $v_{6,\textsc{RP}}$. In the event-by-event hydro simulation, we found that $v_{6,\textsc{RP}}$ is within $\mathcal{O}(10^{-4})$, which makes $\Delta\gamma_3^\mathrm{bkg}\sim\mathcal{O}(10^{-8})$. 
As has been found in the simulation of CME, resonance decays contribute to about $\sim 50\%$ of the non-CME background. Therefore, although some other possible backgrounds, e.g. the local charge conservation, are not included in the current estimation, we expect the overall background to be of the same order as what is shown here.
Hence, we predict that the signal $\Delta\gamma_3^\mathrm{sgn}\sim\mathcal{O}(10^{-7})$  may be  significantly (by an order of magnitude) greater than the background. 
\vskip0.3cm

{\it Discussion - } While the magnitude of the observable $\Delta\gamma_3^\mathrm{sgn}\sim\mathcal{O}(10^{-7})$ induced by siCME and siCVE is about two   orders of magnitude smaller than for ``conventional" CME and CVE, it is expected to be much less contaminated by the background. This is because the resonance decays and local charge conservation contribute a lot less to this observable than to the ``conventional" $\gamma$ correlator.

The expected dominance of the signal over the background (about an order of magnitude) should make siCME and siCVE detectable in heavy ion collisions with high statistics data samples. We thus urge the experimental studies of siCME and siCVE in both AuAu and isobar collisions at RHIC  (even though we do not predict an observable {\it difference} between the two isobar pairs, the effects could be detectable in both isobars with the presently accumulated statistics). It will also be of interest to investigate the effect at the LHC.

It would be interesting to search for siCME and siCVE at lower collision energies during the beam energy scan. One may expect the enhancement of these effects due to the larger baryon chemical potential, possibly larger vorticity~\cite{Singh:2021yba}, longer-lived magnetic field, and the enhancement of topological fluctuations~\cite{Ikeda:2020agk} due to proximity to the critical point of the QCD phase diagram~\cite{Stephanov:1998dy,An:2021wof}.  

In the future, it will be interesting to investigate the contribution, analogous to siCME and siCVE, of anomalous shear-induced axial currents to the polarization of  
 $\Lambda$ hyperons. In particular, the proportionality of the corresponding transport coefficients to the square of the chemical potential $\mu^2$ can yield a characteristic dependence of polarization on  the charge asymmetry of the event. It would also be important to check our predictions for the siCME and siCVE transport coefficients using lattice QCD and the Kubo relations derived here.
 
\vspace*{0.3cm}
\acknowledgments
This work is supported by the U.S. Department of Energy, Office of
Science, Office of Nuclear Physics, Grants Nos. DE-FG88ER41450 and DE-SC0012704 (D.K., S.S., Y.-C.L.), DE-FG02-87ER40371 (M.B.), DE-FG02-92ER40713 (S.V.), DE-FG0201ER41195 (H.-U.Y.),  and
within the framework of the Beam Energy Scan Theory (BEST) Topical
Collaboration.

%

\cleardoublepage
\newpage
\begin{widetext}
\begin{center}
\textbf{\large Supplementary Materials}
\end{center}
\section{Computing siCVE and siCME coefficients using the method of moment expansion}\label{app:1}
Here we show details of computing shear-induced chiral vortical and magnetic effect transport coefficients using the method of moment expansion.
We substitute the distribution function in the current~\eqref{eq:ckt_current} by the non-equilibrium one,
\begin{equation}
\begin{split}
f^\pm =\,& f_\mathrm{eq}^\pm + f_\mathrm{eq}^\pm (1-f_\mathrm{eq}^\pm) 
	\Big( 
    \lambda_{\Pi}^\pm \Pi
		+	\lambda_{\nu}^\pm \nu_{\pm}^\mu p_\mu
		+	\lambda_{\pi}^\pm \pi^{\mu\nu} p_\mu p_\nu 
	\Big).
\end{split}
\end{equation}
where $\lambda_{X}^{\pm}$ are polynomials of $u\cdot p$ (the "comoving energy"), with coefficients being functions of $T$ and $\alpha^\pm$ which are determined by matching the energy-momentum stress tensor from its microscopic integration representation to the corresponding macroscopic viscous terms.
For a Weyl fermion of chirality $\chi=\pm 1$ , the shear-induced chiral vortical and magnetic currents read
\begin{align}
J^{\mu}_\chi =\;&   
    \cdots + \chi \hbar\int_p \frac{\epsilon^{\mu\nu\rho\sigma}p_\rho u_\sigma}{2 u \cdot p} 
    \nabla_\nu f_\chi \\
\begin{split}
=\;&
    \cdots +
      \chi \hbar  \pi^{\alpha\beta} \int_p \bigg[ 
    \frac{\epsilon^{\mu\nu\rho\sigma} p_\rho u_\sigma}{2 u \cdot p}
    (\partial_\nu - Q\, F_{\nu\lambda} \partial_p^\lambda) (f_0^\chi (1-f_0^\chi) \lambda_{\pi}^\chi p_{\langle\alpha} p_{\beta\rangle} )
    \bigg]
\end{split}\\
\begin{split}
=\;&
    \cdots +  \chi \hbar \pi^{\alpha\beta} \epsilon^{\mu\nu\rho\sigma} \bigg[
       Q\, F_{\beta\nu} \int_p
      \frac{p_\rho  u_\sigma}{u \cdot p}
    f_0^\chi (1-f_0^\chi) \lambda_{\pi}^\chi p_{\alpha}
    +(\partial_\nu \beta_\lambda)
     \int_p  \frac{p^\lambda  p_\rho  p_{\langle\alpha} p_{\beta \rangle}
    u_\sigma}{2 u \cdot p} 
    [f_0^\chi (1-f_0^\chi) \lambda_{\pi}^\chi ]'\bigg]
\end{split}\\
\begin{split}
=\;&\cdots
    +  \frac{\chi \hbar}{15} \frac{\epsilon^{\lambda\nu\rho\sigma} {\pi_{\rho}}^{\mu}}{2} u_\sigma (\partial_\nu \beta_\lambda)
     \int_p E^3_p
    [f_0^\chi (1-f_0^\chi) \lambda_{\pi}^\chi ]'
    +
      \frac{\chi \hbar  Q}{3}\,\epsilon^{\mu\rho\sigma\nu} \epsilon_{\nu\lambda\alpha\beta}
      u_\sigma {\pi_{\rho}}^{\lambda} u^\alpha B^\beta \int_p E_p
    f_0^\chi (1-f_0^\chi) \lambda_{\pi}^\chi
\end{split}\\
\begin{split}
=\;&\cdots
    -  \frac{\chi \hbar}{5} {\pi^{\mu}}_{\rho} \omega^{\rho} 
     \int_p E^2_p
    f_0^\chi (1-f_0^\chi) \lambda_{\pi}^\chi
    -
       \frac{\chi \hbar Q}{3}
      {\pi^{\mu}}_{\rho} B^\rho \int_p E_p
    f_0^\chi (1-f_0^\chi) \lambda_{\pi}^\chi,
\end{split}
\end{align}
where $(\cdots)$ refers to the terms that are orthogonal to the siCVE or siCME currents, and $\int_p \equiv \int \frac{\mathrm{d}^4p}{4\pi^3} \delta(p^2)$. In deriving the above, we have used the Shouten identity, the decomposition of the field strengh tensor in terms of electric and magnetic fields, i.e.,
$F_{\mu\nu} = E_\mu u_\nu - E_\nu u_\mu + \epsilon_{\mu\nu\rho\sigma} u^\rho B^\sigma$,
and the identity
$
\epsilon^{\mu\rho\sigma\nu} \epsilon_{\nu\lambda\alpha\beta}=
      \delta^\mu_\lambda \delta^\rho_\alpha \delta^\sigma_\beta
      +\delta^\mu_\alpha \delta^\rho_\beta \delta^\sigma_\lambda
      +\delta^\mu_\beta \delta^\rho_\lambda \delta^\sigma_\alpha
      -\delta^\mu_\lambda \delta^\rho_\beta \delta^\sigma_\alpha
      -\delta^\mu_\alpha \delta^\rho_\lambda \delta^\sigma_\beta
      -\delta^\mu_\beta \delta^\rho_\alpha \delta^\sigma_\lambda
$.

In leading order in the moment expansion, which is usually referred to as the 14-moment formalism for single component system, $\lambda_{\pi}$ is truncated at the zeroth order in comoving energy, i.e., $\lambda_{\pi} = c_0 (T,\mu_\pm) \times (u\cdot p)^0$. We will use this approximation in our computation, which gives us
\begin{align}
\begin{split}
T^{\mu\nu} =\;& 
    \sum_{i}\sum_{\chi=\pm} \int \frac{\mathrm{d}^4p}{4\pi^3}\delta(p^2) p^\mu p^\nu f_{i,\chi} 
+
     \sum_{j} \int \frac{\mathrm{d}^4p}{4\pi^3}\delta(p^2) p^\mu p^\nu f_{j}
     + \cdots
\end{split}\\
\begin{split}
=\;&
	\frac{2 \pi^{\alpha\beta}}{15}  \Bigg[ \sum_{i,\chi} \lambda_{\pi}(\alpha_i^\chi) \int\frac{\mathrm{d}^4p}{4\pi^3}\delta(p^2) E_p^4 f_{i}^\chi (1-f_{i}^\chi) 
	+\sum_{j} \lambda_{\pi}(0) \int\frac{\mathrm{d}^4p}{4\pi^3}\delta(p^2) E_p^4 f_{j} (1+f_{j}) \Bigg] + \cdots,
\end{split}
\end{align}
where $(\cdots)$ represents the terms orthogonal to $\pi^{\mu\nu}$,  $i$($j$) are sums over all the quark(gluon) degrees of freedom. We also assume that the gluon distribution has the same coefficient $\lambda_\pi$ as the quarks at zero chemical potential, for simplicity.
Imposing the matching condition that the above integral should exactly be $\pi^{\alpha\beta}$, we find that
\begin{align}
\begin{split} 
1=\;&
	 N_c N_F \frac{2}{15}\int \frac{\mathrm{d}^4p}{4\pi^3}\delta(p^2) E_p^4 f_{i,\mathrm{eq}}^+ (1-f_{i,\mathrm{eq}}^+) \lambda_{\pi}^+
	 +N_c N_F \frac{2}{15}\int \frac{\mathrm{d}^4p}{4\pi^3}\delta(p^2) E_p^4 f_{i,\mathrm{eq}}^- (1-f_{i,\mathrm{eq}}^-) \lambda_{\pi}^-
\\&
	+(N_c^2-1) \frac{2}{15}\int \frac{\mathrm{d}^4p}{4\pi^3}\delta(p^2) E_p^4 f_B (1+f_B) \lambda_{\pi}
\end{split}\\
\begin{split} 
=\;&
	 \frac{T^6}{\pi^2} \Big[
	 124\,\zeta(5) \lambda_\pi(0)
    + [135\,\zeta(5) + 54\,\zeta(3)\alpha_+^2] \lambda_\pi(\alpha_+)
	+[135\,\zeta(5) +  54\,\zeta(3)\alpha_-^2] \lambda_\pi(\alpha_-)
	\Big] + \mathcal{O}(\alpha^3),
\end{split}
\end{align}
where $N_c=3$ and $N_F=3$ are respectively the numbers of colors and flavors. Therefore,
\begin{align}
    \lambda_\pi(\alpha) =
    \frac{\pi^2/T^6}{394\,\zeta(5)}\Big(1 - \frac{3\,\zeta(3)}{5\,\zeta(5)}\alpha^2 +\mathcal{O}(\alpha^3)\Big).
\end{align}
Using this value in the expressions for the siCME and siCVE, we find
\begin{align}
\begin{split}
    \xi_{1,i,\pm}
=\;& \mp 
    \frac{\hbar \eta}{5} \lambda_{\pi}(\alpha_i^{\pm})
     \int \frac{\mathrm{d}^4p}{4\pi^3}\delta(p^2) E^2_p
    f_{i,\mathrm{eq}}^\pm (1-f_{i,\mathrm{eq}}^\pm)
= \mp
    \frac{\hbar \eta}{T^2} \frac{9\,\zeta(3)}{3940\,\zeta(5)} \big(1 - 0.311\,\alpha_{i,\pm}^2\big) +\mathcal{O}(\alpha^3)\,,
\end{split}
\end{align}
and
\begin{align}
\begin{split}
    \xi_{2,i,\pm}
=\;& \mp 
    \frac{\hbar \eta}{3} \lambda_{\pi}(\alpha_i^{\pm})
     \int \frac{\mathrm{d}^4p}{4\pi^3}\delta(p^2) E_p
    f_{i,\mathrm{eq}}^\pm (1-f_{i,\mathrm{eq}}^\pm)
= \pm
    \frac{\hbar \eta}{T^3} \frac{3\,\zeta(3)}{197\,\zeta(5)} \alpha_{i,\pm} +\mathcal{O}(\alpha^3)\,.
\end{split}
\end{align}
Although we can use the known result for the shear viscosity at leading order  $\eta\sim 1/(\alpha_s^2\log(1/\alpha_s))$ at this point, we instead use the relation, $\eta \approx \frac{s}{4\pi} = 52\times\frac{T^3}{\pi^3}$, which is motivated by experiments.  We then finally arrive at the expressions
\begin{align}
    \xi_{1} =\;& N_c N_F \times(\xi_{1,i,+}+\xi_{1,i,-}) \approx -0.05 \frac{\mu_A \mu}{T},\\
    \xi_{2} =\;& N_c N_F \times(\xi_{2,i,+}+\xi_{2,i,-}) \approx -0.53 \frac{\mu_A}{T}.
\end{align}
We note that it is also possible to go beyond the moment expansion approximation we use here, and obtain the distribution function, and hence $\xi_{1,2}$, more precisely from the full QCD collision terms at leading order in coupling constant. Our computation should be a reasonable approximation to that numerically.

\section{Kubo formula for \texorpdfstring{$\xi_1$}{xi1}}\label{app:2}
We use the Zubarev non-equilibrium statistical operator
method~\cite{Zubarev:1966,Zubarev:1979,vanWeert1982,Zubarev:1989su,Morzov:1998}
to derive the Kubo formula for the siCVE
conductivity. This is the extension to the second order of
the argument used in~\cite{Hosoya:1983id} to derive the Kubo formulas
for transport coefficients, like the shear viscosity, in quantum field
theory. For recent reviews and recent applications of the method
see~\cite{Huang:2011dc,Becattini:2019dxo,Buzzegoli:2020ycf,Harutyunyan:2018cmm,Harutyunyan:2021rmb}.

Denoting with $\Sigma(\tau)$ the 3D space-like hyper-surface of the space-time
foliation parameterized by the ``time'' $\tau$, the covariant statistical operator
of a system which reached the local thermal equilibrium at the ``time'' $\tau_0$ is
obtained by maximizing the total entropy $S=-\tr(\h{\rho}\log\h{\rho})$ with constrained
values of energy–momentum and charge density, which should be equal to the actual
values. The resulting stationary non-equilibrium statistical operator is
\begin{equation*}
\begin{split}
\h{\rho}(\tau_0) =& \frac{1}{Z} \exp\left\{
	-\int_{\Sigma(\tau_0)}\D \Sigma_\mu \left(\h{T}^{\mu\nu}\beta_\nu - \h{j}^\mu \zeta \right)\right\},
\end{split}
\end{equation*}
where $\h{T}$ is the symmetric Belinfante stress-energy tensor.
Taking into account that $\h{T}$ and $\h{j}$ are conserved and using the Gauss theorem, $\h{\rho}(\tau_0)$
can be rewritten in terms of the operators at present ``time'' $\tau$:
\begin{equation}
\label{eq:StatOperFull}
\begin{split}
\h{\rho} =& \frac{1}{Z} \exp\left\{
	-\int_{\Sigma(\tau)}\D \Sigma_\mu \left(\h{T}^{\mu\nu}\beta_\nu - \h{j}^\mu \zeta \right)
	+ \int_\Omega\D\Omega \left(\h{T}^{\mu\nu}\nabla_\mu\beta_\nu - \h{j}^\mu \nabla_\mu \zeta \right)
\right\},
\end{split}
\end{equation}
with $\Omega$ the region of space-time enclosed by the two hyper-surfaces
$\Sigma(\tau_0)$ and $\Sigma(\tau)$ and the time-like hyper-surface at their boundaries.

If we want to evaluate the thermal average of an operator $\h{O}$ around a point $x$
for a system in the hydrodynamic regime we can approximate the statistical operator
$\h\rho$ as
\begin{equation*}
\h{\rho}\simeq \frac{1}{Z} \exp\left\{ \h{A} +\h{B}_\varpi + \h{B}_{\rm D} \right\}
\end{equation*}
where
\begin{equation*}
\begin{split}
\h{A} = & -\beta(x)\cdot \h{P} + \zeta(x) \h{Q},\quad
\h{B}_\varpi =  \frac{1}{2}\varpi_{\rho\sigma}(x)\h{J}^{\rho\sigma}_x, \quad
\h{B}_{\rm D} = \int_\Omega\D\Omega \left(\h{T}^{\mu\nu}\nabla_\mu\beta_\nu - \h{j}^\mu \nabla_\mu \zeta \right),
\end{split}
\end{equation*}
with $\h{P}$ is the total four-momentum of the system and $\h{J}^{\rho\sigma}_x=\h{\group{T}}(x)\h{J}^{\rho\sigma}\h{\group{T}}^\dagger(x)$
are the generators of Lorentz transformations translated by $x$, that is
\begin{equation*}
\begin{split}
\h{J}_x^{\rho\sigma} = \int_{T\Sigma(\tau)} \!\!\!\D^3 x_1 n_\lambda
	\left[ (x_1-x)^\rho \h{T}^{\lambda\sigma} - (x_1-x)^\sigma\h{T}^{\lambda\rho}  \right],
\end{split}
\end{equation*}
and $\h{Q}$ is a conserved charge. Since the siCVE can only occur if the parity
symmetry of the statistical operator is broken we should
also include an axial charge $\h{Q}_A$ inside the operator $\h{A}$. However, the derivation of
the Kubo formula for $\xi_1$ does not relay on the presence of the axial charge and for the
sake of clarity we can introduce it at end. We denoted with $\varpi$ the thermal vorticity
\begin{equation*}
\varpi_{\mu\nu}(x)= -\frac{1}{2}\left(\pd_\mu\beta_\nu(x) - \pd_\nu\beta_\mu(x)\right)
\end{equation*}
and we are neglecting the non-dissipative effects of thermal-shear.
The operator $\h{A}$ reproduces the statistical operator of homogeneous
thermal equilibrium with four-temperature $\beta(x)$
\begin{equation}
\label{eq:StatOperBeta}
\h{\rho}_{\beta}=\frac{1}{Z_{\beta}}\E^{\h{A}},
\end{equation}
and for any operator $\h{O}(x')$ such that $[\h{O}(x'),\h{Q}]=0$ the operator
$\exp(\h{A})$ acts as an imaginary translation along $\beta(x)$:
\begin{equation*}
\begin{split}
\E^{-\lambda \h{A}}\h{O}(x')\E^{\lambda \h{A}} =& \E^{\lambda(\beta(x)\cdot \h{P} - \zeta(x) \h{Q})} \h{O}(x') \E^{-\lambda(\beta(x)\cdot \h{P} - \zeta(x) \h{Q})}
= \h{\group{T}}(-\I \lambda \beta(x))\h{O}(x')\h{\group{T}}^\dagger(-\I \lambda \beta(x))
= \h{O}(x' -\I \lambda \beta(x)).
\end{split}
\end{equation*}
The operator $\h{B}_{\rm D}$ contains the dissipative effects and the
operator $\h{B}_\varpi$ contains the non-dissipative effects of thermal vorticity.

Calling $\h{B}\equiv \h{B}_{\rm D} + \h{B}_\varpi$ we evaluate the thermal
average of an operator $\h{O}(x)$ with linear response theory at second
order of $\h{B}$. We obtain~\cite{vanWeert1982,Buzzegoli:2018wpy}
\begin{equation}
\label{eq:MeanO}
\begin{split}
\mean{\h{O}(x)} =& \tr \left[\h{\rho}\; \h{O}(x) \right]
\simeq \frac{\tr \left[\exp\left\{\h{A}+\h{B}\right\} \h{O}(x) \right]}{\tr \left[\exp\left\{\h{A}+\h{B}\right\} \right]}
\simeq \mean{\h{O}(x)}_\beta + \mean{\h{B}_1 \h{O}(x)}_{\beta,{\rm c}} +\frac{1}{2}\mean{\h{B}_2 \h{O}(x)}_{\beta,{\rm c}}+\cdots ,
\end{split}
\end{equation}
where $\mean{\cdots}_\beta$ denotes the thermal average with the statistical operator in Eq.~(\ref{eq:StatOperBeta}),
the subscript ``c'' denotes the following connected correlators:
\begin{equation*}
\begin{split}
\mean{\h{B}_1 \h{O}(x)}_{\beta,{\rm c}} = & \mean{\h{B}_1 \h{O}(x)}_\beta - \mean{\h{B}_1}_\beta \mean{\h{O}(x)}_\beta,\\
\mean{\h{B}_2 \h{O}(x)}_{\beta,{\rm c}} = & 2 \mean{\h{B}_2 \h{O}(x)}_\beta - 2\mean{\h{B}_2}_\beta \mean{\h{O}(x)}_\beta
	-2\mean{\h{B}_1}_\beta \mean{\h{B}_1\h{O}(x)}_\beta + 2\mean{\h{B}_1}_\beta^2 \mean{\h{O}(x)}_\beta
\end{split}
\end{equation*}
and we defined
\begin{equation}
\label{eq:DefBn}
\begin{split}
\h{B}_1 = & \int_0^1 \D \lambda \h{B}(\lambda) ,\quad
\h{B}_2 = \int_0^1 \D \lambda_1 \int_0^{\lambda_1} \D \lambda_2 \h{B}(\lambda_1)\h{B}(\lambda_2),\quad
\h{B}(\lambda) = \E^{-\lambda \h{A}} \h{B} \E^{\lambda \h{A}}.
\end{split}
\end{equation}
Using
\begin{equation*}
\h{B}_2 = \frac{1}{2} \int_0^1 \D \lambda_1 \int_0^1 \D \lambda_2 T_{\lambda} \left\{\h{B}(\lambda_1)\h{B}(\lambda_2)\right\}
\end{equation*}
where $T_\lambda$ is the time-ordering operator with respect to the variables $\lambda_1$
and $\lambda_2$, it is straightforward to see that
\begin{equation}
\label{eq:ConnectedAndTimeOrdered}
\frac{1}{2}\mean{\h{B}_2 \h{O}(x)}_{\beta,{\rm c}} = \left(\h{B},\,\h{B},\,\h{O}(x) \right)
\end{equation}
where we defined the three-point correlation function as
\begin{equation}
\label{eq:threepoint}
\begin{split}
\big(\h{Y},\,\h{Z},\,\h{X}\big) \equiv& \frac{1}{2}\int_0^1\D\lambda_1\int_0^1\D\lambda_2 T_\lambda
	\left\{ \mean{\h{Y}(\lambda_1) \h{Z}(\lambda_2) \h{X} }_\beta\right.
- \mean{\h{Y}(\lambda_1) \h{Z}(\lambda_2)}_\beta \mean{\h{X} }_\beta - \mean{\h{Y}(\lambda_1)}_\beta \mean{\h{Z}(\lambda_2) \h{X} }_\beta\\
&\left. -\mean{\h{Z}(\lambda_2)}_\beta \mean{\h{Y}(\lambda_1) \h{X} }_\beta
	+ 2 \mean{\h{Y}(\lambda_1)}_\beta \mean{\h{Z}(\lambda_2)}_\beta \mean{\h{X} }_\beta \right\}.
\end{split}
\end{equation}
This shows that the mean value in Eq.~(\ref{eq:MeanO}) agrees with the one
given in~\cite{Harutyunyan:2018cmm,Harutyunyan:2021rmb}. It is also straightforward
to find the identities:
\begin{equation}
\label{eq:ThreeCorrIdentities}
\begin{split}
\left(\h{A} + \h{B},\, \h{C},\, \h{O} \right) = & \left(\h{A},\, \h{C},\, \h{O} \right) + \left(\h{B},\, \h{C},\, \h{O} \right)\\
\left(\h{A},\, \h{B},\, \h{O} \right) = & \left(\h{B},\, \h{A},\, \h{O} \right).
\end{split}
\end{equation}
The second order corrections are then given by:
\begin{equation*}
\begin{split}
\left(\h{B},\,\h{B},\,\h{O}(x) \right) &= \left(\h{B}_\varpi,\,\h{B}_\varpi,\,\h{O}(x) \right)
	+2 \left(\h{B}_\varpi,\,\h{B}_{\rm D},\,\h{O}(x) \right) + \left(\h{B}_{\rm D},\,\h{B}_{\rm D},\,\h{O}(x) \right).
\end{split}
\end{equation*}
The first term of the r.h.s gives the corrections at second
order in thermal vorticity and they were studied in~\cite{Becattini:2015nva,Buzzegoli:2017cqy,Buzzegoli:2018wpy}.
The third term contains the second order corrections and were
studied recently with this method in~\cite{Harutyunyan:2021rmb}.
The second term describes the interplay between the thermal vorticity
and the viscous effects and it was not studied before. Our siCVE arises from this term. We denote it
as
\begin{equation}
\label{eq:DeltaVarpiEta}
\begin{split}
\Delta_{\varpi D} O(x) = 2 \left(\h{B}_\varpi,\,\h{B}_{\rm D},\,\h{O}(x) \right) .
\end{split}
\end{equation}

The integration domain in the operator $\h{B}_{\rm D}$ is the region enclosed by the two
hyper-surfaces $\Sigma(\tau)$ and $\Sigma(\tau_0)$ at times $t$ and $t_0$.
We approximate these hyper-surfaces with the space-like hyperplanes with normal vector $n=\hat{\beta}$
that are tangent at the points $x = ( \tau, \sigma )$ and $x_0 = ( \tau_0 , \sigma )$.
In this way the integration is carried out over Minkowski spacetime in the Cartesian coordinates
where $t$ is the time of an observer moving with velocity $n$, and $\vec{x}$ lays in the hyper-planes.
That is we use
\begin{equation*}
\begin{split}
\h{B}_{\rm D} = & \int_\Omega\D\Omega \left(\h{T}^{\mu\nu}\nabla_\mu\beta_\nu - \h{j}^\mu \nabla_\mu \zeta \right)
\to \int_{T\Omega}\D^4 x_2 \left(\h{T}^{\mu\nu}(x_2)\pd_\mu\beta_\nu(x_2) - \h{j}^\mu(x_2) \pd_\mu \zeta(x_2) \right)
\end{split}
\end{equation*}
where $T\Omega$ is the region encompassed by the two hyper-planes.
We can furthermore define an operator $\h{C}_{\rm D}(x_2)$, such that
\begin{equation*}
\begin{split}
\h{B}_{\rm D} = \int_{t_0}^t \D^4 x_2\, \h{C}_{\rm D}(x_2).
\end{split}
\end{equation*}

The operator $\h{J}_x$ is a conserved operator and its value does not depend
on which integration hyper-surfaces is chosen. Using the hyper-plane tangent
to $\Sigma(\tau)$ we have
\begin{equation}
\label{eq:LorentzTrans}
\begin{split}
\h{J}_x^{\rho\sigma} = \int_{T\Sigma(\tau)} \!\!\!\D^3 x_1 n_\lambda
	\left[ (x_1-x)^\rho \h{T}^{\lambda\sigma} - (x_1-x)^\sigma\h{T}^{\lambda\rho}  \right],
\end{split}
\end{equation}
where $\h{T}^{\lambda\sigma}$ is the symmetric Belinfante stress-energy tensor.
Notice that the time component, i.e. the direction orthogonal to the tangent plane of $\Sigma$,
is directed along $\beta$.
We can also write $\h{B}_\varpi$ in terms of the angular momentum
and boost operators of the system. Using
\begin{equation*}
\frac{1}{2}\varpi:\h{J}_x = - \beta\omega_\rho \h{J}_x^\rho - \beta A_\rho \h{K}_x^\rho,
\end{equation*}
where $\beta = \sqrt{\beta^2}$, $\h{J}$ and $\h{K}$ are the generators of
Lorentz rotation and boost transformations
\begin{equation}
\label{eq:BoostandRotationDef}
\h{K}^\rho = u_\lambda \h{J}^{\lambda\rho},\quad
\h{J}^{\rho} = \frac{1}{2}\epsilon^{\alpha\beta\gamma\rho} u_\alpha \h{J}_{\beta\gamma},
\end{equation}
and
\begin{equation*}
\begin{split}
\omega_\rho = -\frac{1}{2}\epsilon_{\rho\mu\nu\lambda} u^\lambda \pd^\mu u^\nu, \quad
A_\rho = u_\lambda \pd^\lambda u^\rho
\end{split}
\end{equation*}
are the rotation and acceleration of the fluid.
The viscous-vortical effects can then be written as
\begin{equation}
\label{eq:DeltaRotandAcc}
\begin{split}
\Delta_{\varpi D} O&(x) = -2 \beta(x) \omega_\rho(x)\int_{t_0}^t \D^4 x_2\,
	\left(\h{J}_x^{\rho},\,\h{C}_{\rm D}(x_2),\,\h{O}(x) \right)
 -2 \beta(x) A_\rho(x)\int_{t_0}^t \D^4 x_2\,
	\left(\h{K}_x^{\rho},\,\h{C}_{\rm D}(x_2),\,\h{O}(x) \right) .
\end{split}
\end{equation}

To identify the components of these second order corrections
we also decompose the stress-energy tensor and the current operators as
\begin{equation*}
\begin{split}
\h{T}^{\mu\nu} =& \hat{\epsilon} u^{\mu}u^{\nu} - \hat{p}\Delta^{\mu\nu}
	+ \hat{q}^{\mu}u^{\nu} + \hat{q}^{\nu}u^{\mu} + \hat{\pi}^{\mu\nu},\\
\h{j}^{\mu} =& \hat{n} u^\mu +\hat{r}^{\mu},
\end{split}
\end{equation*}
where $\beta^\mu =\beta u^\mu$, $\Delta^{\mu\nu}=\eta^{\mu\nu} - u^\mu u^\nu$
and the quantities are defined such that
\begin{equation*}
u_{\nu}\hat{q}^{\nu}= u_{\nu}\hat{r}^{\nu}=u_{\nu}\hat{\pi}^{\mu\nu}
	=\hat{\pi}_{\mu}^\mu=0.
\end{equation*}
Consequently the above operators are obtained with
\begin{equation*}
\begin{split}
\hat{\energy} =& u_\mu u_\nu \h{T}^{\mu\nu},\,
	\hat{n} = u_\mu\h{j}^{\mu},\,
	\hat{p}=-\frac{1}{3}\Delta_{\mu\nu} \h{T}^{\mu\nu},\\
\hat{\pi}^{\mu\nu} =& \Delta_{\alpha\beta}^{\mu\nu} \h{T}^{\alpha\beta},\,
	\hat{q}^\mu  = u_\alpha\Delta_{\beta}^{\mu}\h{T}^{\alpha\beta},\,
	\hat{r}^{\nu}=\Delta_{\mu}^{\nu} \h{j}^{\mu},
\end{split}
\end{equation*}
where we defined
\begin{equation*}
\Delta_{\mu\nu\rho\sigma}= \frac{1}{2}\left(\Delta_{\mu\rho}\Delta_{\nu\sigma}
	+\Delta_{\mu\sigma}\Delta_{\nu\rho}\right)
	-\frac{1}{3}\Delta_{\mu\nu}\Delta_{\rho\sigma}.
\end{equation*}
In terms of the operators defined above, the operator $\h{C}_{\rm D}$
can be decomposed as follows~\cite{Zubarev:1979,Hosoya:1983id,Huang:2011dc,Harutyunyan:2021rmb}
\begin{equation*}
\begin{split}
\h{C}_{\rm D} = & \h{T}^{\mu\nu}\pd_\mu\beta_\nu - \h{j}^\mu \pd_\mu \zeta
	=  \hat{\energy} D\beta - \hat{p}\beta\theta -\hat{n} D\zeta
	 + \hat{q}_{\sigma}(\beta Du^{\sigma}+\pd^{\sigma}\beta)-\hat{r}_{\sigma} \pd^\sigma\zeta
	 +\beta\hat{\pi}^{\alpha\tau}\sigma_{\alpha\tau},
\end{split}
\end{equation*}
where $D=u^\mu \pd_\mu$, $\theta=\pd_\mu u^\mu$ and $\sigma_{\alpha\tau}$
is the shear-viscosity tensor
\begin{equation*}
\sigma_{\alpha\tau} = \Delta_{\alpha\tau}^{\mu\nu}\pd_\mu u_\nu .
\end{equation*}
For the siCVE we are interested in the shear-viscosity part, that
we denote with
\begin{equation*}
\begin{split}
\h{C}_\eta(x_2) = & \beta(x_2)\sigma_{\alpha\tau}(x_2)\,\hat{\pi}^{\alpha\tau}(x_2),
\end{split}
\end{equation*}
and in the rotation term of (\ref{eq:DeltaRotandAcc}).
The linear response of the shear-rotation coupling will then be generated by
\begin{equation}
\label{eq:DeltaOmegaEtaO}
\begin{split}
\Delta_{\omega\eta} O(x) =& -2 \beta(x) \omega_\rho(x)\int_{t_0}^t \D^4 x_2\,
	\left(\h{J}^{\rho}_x,\,\h{C}_{\eta}(x_2),\,\h{O}(x) \right).
\end{split}
\end{equation}
In particular, for a vector current we have
\begin{equation}
\label{eq:DeltaShearvorticityx2}
\begin{split}
\Delta_{\omega\eta} j^\mu(x) =& -2 \beta(x) \omega_\rho(x)\int_{t_0}^t \D^4 x_2\,
	\left(\h{J}^{\rho}_x,\,\h{C}_{\eta}(x_2),\,\h{j}^\mu(x) \right)\\
	= & -2\beta(x) \omega_\rho(x)\int_{t_0}^t \D^4 x_2\,  \beta(x_2)\sigma_{\alpha\tau}(x_2)
	\left(\h{J}^{\rho}_x,\,\h{\pi}^{\alpha\tau}(x_2),\,\h{j}^\mu(x) \right).
\end{split}
\end{equation}
The properties under discrete transformation of the operators
in the above correlator are
\[\begin{array}{lccc}
	& \h{\vec{J}} & \h{\pi}^{ij}=\h{T}^{ij} & \h{\vec{j}}\\  
	\hline
	\group{P} & + & + & - \\
	\group{T} & - & + & - \\
	\group{C} & + & + & - \\
\end{array}\]
We see that the time reversal symmetry is preserved but
a breaking of parity (and charge) symmetry in the
statistical operator is needed to obtain a non-vanishing result,
hence this effect is chiral. We also learn that the acceleration
term in Eq.~(\ref{eq:DeltaRotandAcc}) does not give a
similar effect because the boost operator has opposite
time-reversal symmetry compared to the angular momentum.

We can now take advantage of the Curie principle~\cite{Zubarev:1979,Hosoya:1983id}
to identify the transport coefficient $\xi_1$.
Since the statistical operator $\h{\rho}_\beta$ is symmetric under SO(3) rotations,
we can only have non vanishing results in the thermal correlators only when averaging
two operators in the same irreducible representation under rotation.
In our case we find, see~\cite[eq. (152)]{Harutyunyan:2018cmm}
\begin{equation}
\label{eq:CuriePrinciple}
\left(\h{J}^{\rho}_x,\,\h{\pi}^{\alpha\tau},\,\h{j}^\mu(x) \right) =
	\frac{\Delta^{\alpha\tau\rho\mu}(x)}{5} \left(\h{J}^{\delta}_x,\,\h{\pi}_{\delta\gamma},\,\h{j}^\gamma(x) \right).
\end{equation}
Using Eq. (\ref{eq:CuriePrinciple}) and
\[
	\omega_\rho \sigma_{\alpha\tau}\Delta^{\alpha\tau\rho\mu}=\sigma^{\mu\nu}\omega_\nu
\]
in Eq.~(\ref{eq:DeltaShearvorticityx2}) we obtain:
\begin{equation*}
\begin{split}
\Delta_{\omega\eta} j^\mu(x) =&-\frac{2}{5}\beta(x) \omega_\nu(x)
    \int_{t_0}^t \D^4 x_2\,  \beta(x_2)\sigma^{\mu\nu}(x_2)
	\left(\h{J}^{\delta}_x,\,\h{\pi}_{\delta\gamma}(x_2),\,\h{j}^\gamma(x) \right).
\end{split}
\end{equation*}
Notice that three-point correlator is now a Lorentz scalar. This means it can be evaluated in any
reference frame. It is convenient to go into the local rest frame where the fluid
velocity is $u(x)=(1,\vec{0})$. In that frame the statistical operator
$\h{\rho}_\beta$ simply becomes the statistical operator
\begin{equation}
\label{eq:StatOperT}
\h{\rho}_T = \frac{1}{Z_T}\exp\left\{-\frac{1}{T}\h{H} + \frac{\mu}{T}\h{Q}
     + \frac{\mu_A}{T}\h{Q}_A \right\},
\end{equation}
where we also introduced an axial charge.
Reminding that $\h{\pi}_{\mu\nu}=\Delta_{\mu\nu\rho\sigma}\h{T}^{\rho\sigma}$
and that $\Delta_{\mu\nu}^{\hphantom{\mu\nu}\mu\nu}=5$ we have
\begin{equation*}
\begin{split}
 \frac{1}{5}\left(\h{J}^\mu,\,\h{\pi}_{\mu\nu},\,\h{j}^\nu \right)
	=  \frac{1}{5}\Delta_{\mu\nu\rho\sigma}\left(\h{J}^\mu,\,\h{T}_{\rho\sigma},\,\h{j}^\nu \right)
	=  \frac{1}{5}\Delta_{\mu\nu\rho\sigma}\eta^{\mu\rho}\eta^{\nu\sigma}\left(\h{J}^i,\,\h{T}^{ij},\,\h{j}^j \right)_{i\neq j}
	= \left(\h{J}_x^x,\,\h{T}^{xy},\,\h{j}^y \right),
\end{split}
\end{equation*}
where we used the Curie principle again. The anomalous current
reads:
\begin{equation*}
\begin{split}
\Delta_{\omega\eta} j^\mu(x) =&-2\beta(x) \omega_\nu(x)
    \int_{t_0}^t \D^4 x_2\,  \beta(x_2)\sigma^{\mu\nu}(x_2)
	\left(\h{J}_x^x,\,\h{T}^{xy}(x_2),\,\h{j}^y(x) \right)_T,
\end{split}
\end{equation*}
where the subscript $T$ means that the thermal averages in the three-point correlator defined in
Eq. (\ref{eq:threepoint}) must be taken with the statistical operator (\ref{eq:StatOperT}).
Such thermal averages can be evaluated with standard techniques of the finite temperature field theory.
Using the definition of Lorentz transformations (\ref{eq:LorentzTrans}) and of rotations
(\ref{eq:BoostandRotationDef}) in the local rest frame we have
\begin{equation}
\label{eq:DeltaShearvorticityCurve}
\begin{split}
\Delta_{\omega\eta} j^\mu(x) =& - 4 \beta(x) \omega_\nu(x)
    \int_{t_0}^t \D^4 x_2 \, \int \D^3 x_1  \beta(x_2)\sigma^{\mu\nu}(x_2)
	  (x_1-x)^y \left(\h{T}^{tz}(x_1),\,\h{T}^{xy}(x_2),\,\h{j}^y(x) \right)_T.
\end{split}
\end{equation}

It has been shown~\cite{Grossi:2014} that if we send the initial time
$t_0$ to $-\infty$ and we suppose that correlators between operators evaluated in
the infinitely remote past and at finite time factorize, then
the above expression is the integral of the retarded three-point Green function:
\begin{equation}
\label{eq:DeltaShearvorticity}
\begin{split}
\Delta_{\omega\eta} j^\mu(x) =& \frac{2 \omega_\nu(x)}{\beta(x)}
    \int \D^4 x_1 \int \D^4 x_2 \int_{-\infty}^{t_2}\!\!\!\D\theta_2\, \beta(x_2)\sigma^{\mu\nu}(x_2)\\
	&\times  (x_1-x)^y\, \I\, G^{\rm R1}_{\h{j}^y,\h{T}^{tz},\h{T}^{xy}}\left(x;\, x_1,\, (\theta_2, \vec{x}_2)\right),
\end{split}
\end{equation}
where~\cite{Evans:1990qh,Evans:1991ky}
\begin{equation*}
\begin{split}
G^{\rm R1}_{\h{O},\h{X},\h{Y}}(x;\, x_1,\, x_2) = &
	-\I\theta( t -t_1) \theta( t_1 -t_2)\meanlr{\left[\left[\h{O}(x),\, \h{X}(x_1)\right], \h{Y}(x_2)\right]}_T\\
	&-\I\theta( t -t_2) \theta( t_2 -t_1)\meanlr{\left[\left[\h{O}(x),\,\h{Y}(x_2)\right],  \h{X}(x_1)\right]}_T ;
\end{split}
\end{equation*}
see also Section~\ref{GreenFunction} below.

In Eq.~(\ref{eq:DeltaShearvorticity}) the anomalous current at certain time is
obtained by integrating the whole evolution of the thermodynamic fields $\beta$ and $\sigma$.
However, for a fluid in the hydrodynamic regime we expect that the thermal correlator
in Eq.~(\ref{eq:DeltaShearvorticity}) is suppressed within distances $|x_2-x|$
much shorter than the length in which the temperature and the shear tensor vary.
In spite of that, one can not simply take $\beta$ and $\sigma$ out of the integral.
The reason being that the nonequilibrium statistical operator (\ref{eq:StatOperFull})
requires the vanishing of the flux of $\h{T}^{\mu\nu}\beta_\nu -\h{j}^\mu\zeta$
at the boundary timelike hypersurface. We should instead study perturbations
of the thermodynamics fields with respect to their equilibrium values, that is
\begin{equation*}
\delta\beta = \beta - \beta(x),
\end{equation*}
such that they are vanishing at the boundary and that they keep the flux vanishing,
for instance by enforcing periodicity of the perturbations in $\vec{x} -\vec{x}_2$.
We can then expand the perturbations in Fourier series. In the hydrodynamic limit,
only the components with very small frequency $q_0$ and very small wave vector $\vec{q}$ will
contribute to the integral in (\ref{eq:DeltaShearvorticity}). We can then consider
the perturbation
\begin{equation*}
\delta\beta_\nu(x_2)\simeq A_\nu \frac{1}{2\I}\left[\E^{\I q\cdot (x_2-x)} - \E^{-\I q\cdot(x_2-x)} \right],
\end{equation*}
where $A_\nu$ is a real constant denoting the amplitude of the smallest wave
four-vector Fourier component. In a compact domain, choosing $q^i=\pi/L^i$ ensure
the vanishing of the flux and $\delta\beta(x_2)=0$. Using this perturbation
the thermodynamic fields inside the integral of (\ref{eq:DeltaShearvorticity})
becomes:
\begin{equation*}
\beta(x_2)\sigma^{\mu\nu}(x_2) = \text{Re} \beta(x)\sigma^{\mu\nu}(x)
    \E^{-\I q\cdot (x_2-x)}
\end{equation*}
We can use this expression in Eq. (\ref{eq:DeltaShearvorticity}) taking the
limit of $q \to 0$, which corresponds to the limit of infinite volume
\begin{equation*}
\begin{split}
\Delta_{\omega\eta} j^\mu(x) =& \sigma^{\mu\nu}(x) \omega_\nu(x) \xi_1 \\
	=&\sigma^{\mu\nu}(x) \omega_\nu(x)
 2\text{Re}  \int \D^4 x_2 \int_{-\infty}^{t_2} \D\theta_2 \int_{t_0}^t \D^4 x_1  (x_1-x)^y
   \I G^{\rm R1}_{\h{j}^y,\h{T}^{tz},\h{T}^{xy}}\left(x;\, x_1,\, (\theta_2, \vec{x}_2)\right) \E^{-\I q\cdot(x_2-x)}.
\end{split}
\end{equation*}
The integral in $t_2$ can be done by parts (see Section~\ref{sec:ByParts}) obtaining
\begin{equation*}
\begin{split}
\xi_1& = 2 n^\alpha\left.\frac{\pd}{\pd q^\alpha}\right|_{q\cdot n=0}\lim_{\vec{q}_T\to 0} \text{Im}
    \int \D^4 x_2 \int \D^4 x_1  (x_1-x)^y
   \I G^{\rm R1}_{\h{j}^y,\h{T}^{tz},\h{T}^{xy}}(x;\, x_1,\, x_2) \E^{-\I q\cdot(x_2-x)}\\
    & = 2 n^\alpha\left.\frac{\pd}{\pd q^\alpha}\right|_{q\cdot n=0}\lim_{\vec{q}_T\to 0} \text{Im}
        \int \D^4 x_2 \int \D^4 x_1  (x_1)^y
         \I G^{\rm R1}_{\h{j}^y,\h{T}^{tz},\h{T}^{xy}}(0;\, x_1,\, x_2) \E^{-\I q\cdot x_2},
\end{split}
\end{equation*}
where $q_T$ is the projection of $q$ orthogonal to $n$
and in the last step we took advantage of the translational invariance of the
Green function and renamed the integration variables $x_1$ and $x_2$.

Reminding the momentum space Green function definition
\begin{equation*}
\begin{split}
G^{\rm R1}_{\h{j}^y,\h{T}^{tz},\h{T}^{xy}}(p,q) =& \int \D^4 x_1 \int \D^4 x_2\,
	G^{\rm R1}_{\h{j}^y,\h{T}^{tz},\h{T}^{xy}}(0;\, x_1,\, x_2) \E^{-\I(p\cdot x_1 + q\cdot x_2)},
\end{split}
\end{equation*}
we readily see that the Kubo formula for $\xi_1$ is
\begin{equation}
\label{eq:Xi1KuboFormulaApp}
\xi_1 = \lim_{p,q\to 0} 2 \frac{\pd}{\pd q^0}\frac{\pd}{\pd p^y}
	\text{Im} G^{\rm R1}_{\h{j}^y,\h{T}^{tz},\h{T}^{xy}}(p,q) .
\end{equation}
%

\subsection{Three-point Green function}
\label{GreenFunction}
Here we show that the three-point correlator in Eq. (\ref{eq:DeltaShearvorticityCurve})
\begin{equation}
\label{eq:DeltaOmegaEtaOwithT}
\Delta_{\omega\eta}O(x)= \int_{T\Sigma}\!\D^3 x_1\! \int \D^4 x_2\, n^\lambda (x_1-x)_\rho
	\left(\h{T}_{\lambda\sigma}(x_1),\,\h{T}_{\gamma\delta}(x_2),\,\h{O}(x) \right)\beta(x_2)\sigma^{\mu\nu}(x_2)
\end{equation}
is equal to
\begin{equation}
\label{eq:DeltaOmegaEtaORetardedGreen}
\Delta_{\omega\eta}O(x)=-\frac{1}{2 \beta(x)^2} \int\D^4 x_1 \int\D^4 x_2 \int_{-\infty}^{t_2}\!\D\theta_2\, n^\lambda (x_1-x)_\rho
	\,\I\, G^{\rm R1}_{\h{O},\h{T}_{\lambda\sigma},\h{T}_{\gamma\delta}}\left(x;\,x_1,\,(\theta_2, \vec{x}_2)\right)\beta(x_2)\sigma^{\mu\nu}(x_2)
\end{equation}
where we denoted with $G^{\rm R1}$ the retarded three-point Green function~\cite{Evans:1990qh,Evans:1991ky}
\begin{equation*}
\begin{split}
G^{\rm R1}_{\h{O},\h{X},\h{Y}}(x;\,x_1,\,x_2) = &
	-\I\,\theta\left( t -t_1 \right) \theta\left( t_1 -t_2 \right)\meanlr{\left[\left[\h{O}(x),\, \h{X}(x_1)\right], \h{Y}(x_2)\right]}_\beta\\
	&-\I\,\theta\left( t -t_2 \right) \theta\left( t_2 -t_1 \right)\meanlr{\left[\left[\h{O}(x),\,\h{Y}(x_2)\right],  \h{X}(x_1)\right]}_\beta .
\end{split}
\end{equation*}

First, reminding the Eq.s (\ref{eq:DefBn}) and (\ref{eq:ConnectedAndTimeOrdered})
and using
\begin{equation*}
\E^{-\lambda \h{A}} \h{X}(t,\vec{x}) \E^{\lambda \h{A}} =
	\group{T}(-\I\lambda \beta) \h{X}(t,\vec{x}) \group{T}^\dagger(-\I\lambda \beta) =
	\h{X}(t-\I\lambda\beta,\vec{x}),
\end{equation*}
we define $A$ and $B$ as
\begin{equation}
\label{eq:DeltaOmegaEtaAplusB}
\Delta_{\omega\eta}O(x)= \int_{T\Sigma}\!\D^3 x_1 \!\int\!\D^4 x_2\,\, n^\lambda (x_1-x)_\rho
	\left(A + B \right)\beta(x_2)\sigma^{\mu\nu}(x_2),
\end{equation}
where
\begin{equation*}
\begin{split}
A =& \int_0^1 \!\!\!\D\lambda_1 \int_0^{\lambda_1} \!\!\!\D\lambda_2 \left[
	\mean{\h{T}_{\lambda\sigma}(t_1-\I\lambda_1\beta,\vec{x}_1) \h{T}_{\gamma\delta}(t_2-\I\lambda_2\beta,\vec{x}_2)\h{O}(x)}_\beta
	-\mean{\h{T}_{\lambda\sigma}(t_1-\I\lambda_1\beta,\vec{x}_1) \h{T}_{\gamma\delta}(t_2-\I\lambda_2\beta,\vec{x}_2)}_\beta\mean{\h{O}(x)}_\beta \right], \\
B=& \int_0^1 \!\!\!\D\lambda_1 \int_0^1 \!\!\!\D\lambda_2 \left[
	\mean{\h{T}_{\lambda\sigma}(t_1-\I\lambda_1\beta,\vec{x}_1)}_\beta \mean{\h{T}_{\gamma\delta}(t_2-\I\lambda_2\beta,\vec{x}_2)\h{O}(x)}_\beta
	+\mean{\h{T}_{\gamma\delta}(t_2-\I\lambda_2\beta,\vec{x}_2)}_\beta \mean{\h{T}_{\lambda\sigma}(t_1-\I\lambda_1\beta,\vec{x}_1) \h{O}(x)}_\beta\right.\\
	&\left.-2 \mean{\h{T}_{\lambda\sigma}(t_1-\I\lambda_1\beta,\vec{x}_1)}_\beta \mean{\h{T}_{\gamma\delta}(t_2-\I\lambda_2\beta,\vec{x}_2)}_\beta \mean{\h{O}(x)}_\beta
	\right].
\end{split}
\end{equation*}
Then we take advantage of the symmetry and we write $A$ as the sum of $A_1$ and $A_2$, where
\begin{equation*}
\begin{split}
A_1 =& \frac{1}{2}\!\int_0^1 \!\!\!\D\lambda_1 \!\!\int_0^{\lambda_1} \!\!\!\D\lambda_2 \!\!\left[
	\mean{\h{T}_{\lambda\sigma}(t_1-\I\lambda_1\beta,\vec{x}_1) \h{T}_{\gamma\delta}(t_2-\I\lambda_2\beta,\vec{x}_2)\h{O}(x)}_\beta
	-\mean{\h{T}_{\lambda\sigma}(t_1-\I\lambda_1\beta,\vec{x}_1) \h{T}_{\gamma\delta}(t_2-\I\lambda_2\beta,\vec{x}_2)}_\beta\mean{\h{O}(x)}_\beta \right], \\
A_2 =& \frac{1}{2}\!\int_0^1 \!\!\!\D\lambda_2 \!\!\int_0^{\lambda_2} \!\!\!\D\lambda_1 \!\!\left[
	\mean{\h{T}_{\gamma\delta}(t_2-\I\lambda_2\beta,\vec{x}_2)\h{T}_{\lambda\sigma}(t_1-\I\lambda_1\beta,\vec{x}_1) \h{O}(x)}_\beta
	-\mean{\h{T}_{\gamma\delta}(t_2-\I\lambda_2\beta,\vec{x}_2) \h{T}_{\lambda\sigma}(t_1-\I\lambda_1\beta,\vec{x}_1)}_\beta\mean{\h{O}(x)}_\beta \right].
\end{split}
\end{equation*}
In $A_1$ we replace the operator $\h{T}_{\gamma\delta}(t_2-\I\lambda_2\beta,\vec{x}_2)$ with the same operator
at time $t=t_0$ plus its time evolution:
\begin{equation*}
\begin{split}
\h{T}_{\gamma\delta}(t_2-\I\lambda_2\beta,\vec{x}_2) =& \h{T}_{\gamma\delta}(t_0-\I\lambda_2\beta,\vec{x}_2)
	+ \int_0^{t_2} \D\theta_2 \frac{\pd}{\pd\theta_2}\h{T}_{\gamma\delta}(\theta_2-\I\lambda_2\beta,\vec{x}_2)\\
=& \h{T}_{\gamma\delta}(t_0-\I\lambda_2\beta,\vec{x}_2)
		+ \int_0^{t_2} \D\theta_2 \frac{\I}{\beta}\frac{\pd}{\pd\lambda_2}\h{T}_{\gamma\delta}(\theta_2-\I\lambda_2\beta,\vec{x}_2),
\end{split}
\end{equation*}
where in the last step we replaced the derivative with respect to $\theta_2$ with the derivative of $\lambda_2$.
After this replacement we integrate in $\lambda_2$, and we obtain
\begin{equation*}
\begin{split}
A_1 = & \frac{1}{2}\! \int_0^1 \!\!\D\lambda_1 \!\!\int_0^{\lambda_1} \!\!\D\lambda_2 \!\left[
	\mean{\h{T}_{\lambda\sigma}(t_1-\I\lambda_1\beta,\vec{x}_1) \h{T}_{\gamma\delta}(t_0-\I\lambda_2\beta,\vec{x}_2)\h{O}(x)}_\beta
	-\mean{\h{T}_{\lambda\sigma}(t_1-\I\lambda_1\beta,\vec{x}_1) \h{T}_{\gamma\delta}(t_0-\I\lambda_2\beta,\vec{x}_2)}_\beta\mean{\h{O}(x)}_\beta \right]\\
&+\frac{1}{2} \int_0^1 \!\D\lambda_1 \!\int_{t_0}^{t_2} \!\!\D\theta_2 \,\frac{\I}{\beta} \left[
		\mean{\h{T}_{\lambda\sigma}(t_1-\I\lambda_1\beta,\vec{x}_1) \h{T}_{\gamma\delta}(\theta_2-\I\lambda_1\beta,\vec{x}_2)\h{O}(x)}_\beta
		-\mean{\h{T}_{\lambda\sigma}(t_1-\I\lambda_1\beta,\vec{x}_1) \h{T}_{\gamma\delta}(\theta_2,\vec{x}_2)\h{O}(x)}_\beta\right.\\
		&-\left.\left(\mean{\h{T}_{\lambda\sigma}(t_1-\I\lambda_1\beta,\vec{x}_1) \h{T}_{\gamma\delta}(\theta_2-\I\lambda_1\beta,\vec{x}_2)}_\beta
		-\mean{\h{T}_{\lambda\sigma}(t_1-\I\lambda_1\beta,\vec{x}_1) \h{T}_{\gamma\delta}(\theta_2,\vec{x}_2)}_\beta\right)\mean{\h{O}(x)}_\beta\right]\\
\equiv& A_1^0 + A_1^{\theta_2},
\end{split}
\end{equation*}
with $A_1^0$ the integral in the first line and $A_1^{\theta_2}$ the integral in the last two lines.
In the limit of $t_0$ that goes to $-\infty$, we have
\begin{equation*}
\lim_{t_0\to-\infty} \h{T}_{\gamma\delta}(t_0-\I\lambda_2\beta,\vec{x}_2)
	\simeq \h{T}_{\gamma\delta}(t_0,\vec{x}_2).
\end{equation*}
At an infinitely remote past operators were no longer correlated~\cite{Hosoya:1983id}, that is for instance
\begin{equation*}
\begin{split}
\lim_{t_0\to-\infty}& \mean{\h{T}_{\lambda\sigma}(t_1-\I\lambda_1\beta,\vec{x}_1) \h{T}_{\gamma\delta}(t_0-\I\lambda_2\beta,\vec{x}_2)\h{O}(x)}_\beta
	=\lim_{t_0\to-\infty} \mean{\h{T}_{\lambda\sigma}(t_1-\I\lambda_1\beta,\vec{x}_1) \h{T}_{\gamma\delta}(t_0,\vec{x}_2)\h{O}(x)}_\beta\\
	=& \lim_{t_0\to-\infty} \mean{\h{T}_{\lambda\sigma}(t_1-\I\lambda_1\beta,\vec{x}_1)\h{O}(x)}_\beta \mean{ \h{T}_{\gamma\delta}(t_0,\vec{x}_2)}_\beta
	= \mean{\h{T}_{\lambda\sigma}(t_1-\I\lambda_1\beta,\vec{x}_1)\h{O}(x)}_\beta \mean{ \h{T}_{\gamma\delta}(t_2,\vec{x}_2)}_\beta,
\end{split}
\end{equation*}
where in the last step we used the translational symmetry of the homogeneous equilibrium.
Similarly, for other terms we have
\begin{equation*}
\begin{split}
\lim_{t_0\to-\infty} A_1^0 = & \frac{1}{2} \!\int_0^1 \!\!\!\D\lambda_1 \!\!\int_0^{\lambda_1} \!\!\!\D\lambda_2 \!\!\left[
	\mean{\h{T}_{\lambda\sigma}(t_1-\I\lambda_1\beta,\vec{x}_1) \h{O}(x)}_\beta \mean{\h{T}_{\gamma\delta}(t_2,\vec{x}_2)}_\beta
	-\mean{\h{T}_{\lambda\sigma}(t_1-\I\lambda_1\beta,\vec{x}_1)}_\beta \mean{\h{T}_{\gamma\delta}(t_2,\vec{x}_2)}_\beta\mean{\h{O}(x)}_\beta \right].
\end{split}
\end{equation*}

We now split $A_1^{\theta_2}$ in $A_1^{\theta_2 0}$ and $A_1^{\theta_2 \theta_1}$, where
\begin{equation*}
\begin{split}
A_1^{\theta_2 0}\!\! =& -\frac{1}{2}\!\! \int_0^1\!\!\! \D\lambda_1 \int_{t_0}^{t_2}\!\!\! \D\theta_2
    \frac{\I}{\beta} \left[
	\mean{\h{T}_{\lambda\sigma}(t_1-\I\lambda_1\beta,\vec{x}_1) \h{T}_{\gamma\delta}(\theta_2,\vec{x}_2)\h{O}(x)}_\beta
	-\mean{\h{T}_{\lambda\sigma}(t_1-\I\lambda_1\beta,\vec{x}_1) \h{T}_{\gamma\delta}(\theta_2,\vec{x}_2)}_\beta\mean{\h{O}(x)}_\beta\right],\\
A_1^{\theta_2\theta_1}\!\! =&\frac{1}{2}\!\! \int_0^1 \!\!\!\D\lambda_1\!\!\! \int_{t_0}^{t_2}
    \!\!\!\!\!\!\D\theta_2 \frac{\I}{\beta}\!\! \left[
 	\mean{\h{T}_{\lambda\sigma}(t_1-\I\lambda_1\beta,\vec{x}_1) \h{T}_{\gamma\delta}(\theta_2-\I\lambda_1\beta,\vec{x}_2)\h{O}(x)}_\beta
 	\!-\!\mean{\h{T}_{\lambda\sigma}(t_1-\I\lambda_1\beta,\vec{x}_1) \h{T}_{\gamma\delta}(\theta_2-\I\lambda_1\beta,\vec{x}_2)}_\beta\mean{\h{O}(x)}_\beta \right].
\end{split}
\end{equation*}
In  $A_1^{\theta_2 0}$ the dependence of $\lambda_1$ is only contained in $\h{T}_{\lambda\sigma}$, then, as done before,
we can replace
\begin{equation*}
\begin{split}
\h{T}_{\lambda\sigma}(t_1-\I\lambda_1\beta,\vec{x}_1) =& \h{T}_{\lambda\sigma}(t_0-\I\lambda_1\beta,\vec{x}_1)
		+ \int_{t_0}^{t_1} \D\theta_1 \frac{\I}{\beta}\frac{\pd}{\pd\lambda_1}\h{T}_{\lambda\sigma}(\theta_1-\I\lambda_1\beta,\vec{x}_1),
\end{split}
\end{equation*}
and integrate in $\lambda_1$:
\begin{equation*}
\begin{split}
A_1^{\theta_2 0} = & -\frac{\I}{2\beta}\int_0^1 \D\lambda_1 \int_{t_0}^{t_2}\D\theta_2\left[
	\mean{\h{T}_{\lambda\sigma}(t_0-\I\lambda_1\beta,\vec{x}_1) \h{T}_{\gamma\delta}(\theta_2,\vec{x}_2)\h{O}(x)}_\beta
	-\mean{\h{T}_{\lambda\sigma}(t_0-\I\lambda_1\beta,\vec{x}_1) \h{T}_{\gamma\delta}(\theta_2,\vec{x}_2)}_\beta\mean{\h{O}(x)}_\beta\right]\\
&+\frac{1}{2\beta^2}\int_{t_0}^{t_1} \D\theta_1 \int_{t_0}^{t_2}\D\theta_2\left[
	\mean{\h{T}_{\lambda\sigma}(\theta_1-\I\beta,\vec{x}_1) \h{T}_{\gamma\delta}(\theta_2,\vec{x}_2)\h{O}(x)}_\beta
	-\mean{\h{T}_{\lambda\sigma}(\theta_1,\vec{x}_1) \h{T}_{\gamma\delta}(\theta_2,\vec{x}_2)\h{O}(x)}_\beta\right.\\
&-\left.\left(\mean{\h{T}_{\lambda\sigma}(\theta_1-\I\beta,\vec{x}_1) \h{T}_{\gamma\delta}(\theta_2,\vec{x}_2)}_\beta
	-\mean{\h{T}_{\lambda\sigma}(\theta_1,\vec{x}_1) \h{T}_{\gamma\delta}(\theta_2,\vec{x}_2)}_\beta\right)\mean{\h{O}(x)}_\beta\right].
\end{split}
\end{equation*}
The second and third line of previous equation can be simplified using the Kubo-Martin-Schwinger (KMS) relations
\begin{equation*}
\begin{split}
\mean{\h{A}(t_1-\I\beta)\h{B}(t_2) \h{C}(t_3) }_\beta =& \mean{\h{B}(t_2) \h{C}(t_3) \h{A}(t_1)}_\beta,\\\
\mean{\h{A}(t_1-\I\beta)\h{B}(t_2)}_\beta =& \mean{\h{B}(t_2) \h{A}(t_1)}_\beta.
\end{split}
\end{equation*}
The first line is evaluated in the limit $t_0\to-\infty$ as done before. We obtain:
\begin{equation}
\label{eq:A1Theta20Final}
\begin{split}
\lim_{t_0\to-\infty} A_1^{\theta_2 0} = & -\frac{\I}{2\beta}\int_0^1 \D\lambda_1 \int_{-\infty}^{t_2}\D\theta_2\left[
	\mean{\h{T}_{\lambda\sigma}(x_1)}_\beta \mean{\h{T}_{\gamma\delta}(\theta_2,\vec{x}_2)\h{O}(x)}_\beta
	-\mean{\h{T}_{\lambda\sigma}(x_1)}_\beta \mean{\h{T}_{\gamma\delta}(\theta_2,\vec{x}_2)}_\beta\mean{\h{O}(x)}_\beta\right]\\
&+\frac{1}{2\beta^2}\int_{-\infty}^{t_1} \D\theta_1 \int_{-\infty}^{t_2}\D\theta_2\left[
	\mean{\h{T}_{\gamma\delta}(\theta_2,\vec{x}_2)\h{O}(x)\h{T}_{\lambda\sigma}(\theta_1,\vec{x}_1)}_\beta
	-\mean{\h{T}_{\lambda\sigma}(\theta_1,\vec{x}_1) \h{T}_{\gamma\delta}(\theta_2,\vec{x}_2)\h{O}(x)}_\beta\right.\\
&+\left.\meanlr{\left[\h{T}_{\lambda\sigma}(\theta_1,\vec{x}_1), \h{T}_{\gamma\delta}(\theta_2,\vec{x}_2)\right]}_\beta \mean{\h{O}(x)}_\beta\right].
\end{split}
\end{equation}

Similarly, in $A_1^{\theta_2\theta_1}$ the integration in $\lambda_1$ can be written as:
\begin{equation*}
\begin{split}
A_1^{\theta_2\theta_1} =&-\frac{1}{2\beta^2}\!\! \int_{t_0}^{t_1} \!\!\!\D\theta_1\!\!\! \int_{t_0}^{\theta_1}
    \!\!\!\!\!\!\D\theta_2 \left[
 	\mean{\h{T}_{\lambda\sigma}(\theta_1-\I\beta,\vec{x}_1) \h{T}_{\gamma\delta}(\theta_2-\I\beta,\vec{x}_2)\h{O}(x)}_\beta
	-\mean{\h{T}_{\lambda\sigma}(\theta_1,\vec{x}_1) \h{T}_{\gamma\delta}(\theta_2,\vec{x}_2)\h{O}(x)}_\beta\right.\\
&\left. -\mean{\h{T}_{\lambda\sigma}(\theta_1-\I\beta,\vec{x}_1) \h{T}_{\gamma\delta}(\theta_2-\I\beta,\vec{x}_2)}_\beta\mean{\h{O}(x)}_\beta
	+\mean{\h{T}_{\lambda\sigma}(\theta_1,\vec{x}_1) \h{T}_{\gamma\delta}(\theta_2,\vec{x}_2)}_\beta\mean{\h{O}(x)}_\beta\right].
\end{split}
\end{equation*}
By applying the Kubo-Martin-Schwinger (KMS) relation two times, we obtain
\begin{equation*}
\mean{\h{A}(t_1-\I\beta,\vec{x}_1)\h{B}(t_2-\I\beta,\vec{x}_2)\h{C}(t_3,\vec{x}_3)} = \mean{\h{C}(t_3,\vec{x}_3)\h{A}(t_1,\vec{x}_1)\h{B}(t_2,\vec{x}_2)},
\end{equation*}
and hence
\begin{equation*}
\begin{split}
A_1^{\theta_2\theta_1} =&-\frac{1}{2\beta^2}\!\! \int_{t_0}^{t_1} \!\!\!\D\theta_1\!\!\! \int_{t_0}^{\theta_1}
    \!\!\!\!\!\!\D\theta_2 \left[
 	\mean{\h{O}(x)\h{T}_{\lambda\sigma}(\theta_1,\vec{x}_1) \h{T}_{\gamma\delta}(\theta_2,\vec{x}_2)}_\beta
	-\mean{\h{T}_{\lambda\sigma}(\theta_1,\vec{x}_1) \h{T}_{\gamma\delta}(\theta_2,\vec{x}_2)\h{O}(x)}_\beta\right.\\
&\left. -\mean{\h{T}_{\lambda\sigma}(\theta_1,\vec{x}_1) \h{T}_{\gamma\delta}(\theta_2,\vec{x}_2)}_\beta\mean{\h{O}(x)}_\beta
	+\mean{\h{T}_{\lambda\sigma}(\theta_1,\vec{x}_1) \h{T}_{\gamma\delta}(\theta_2,\vec{x}_2)}_\beta\mean{\h{O}(x)}_\beta\right].
\end{split}
\end{equation*}
In the lint $t_0\to\infty$ we split the quantity $A_1$ into a connected and disconnected part: $A_1=A_1^{C}+A_1^{DC}$, with
\begin{equation*}
\begin{split}
A_1^{C} = & -\frac{1}{2\beta^2}\!\! \int_{-\infty}^{t_1} \!\D\theta_1\!\int_{-\infty}^{\theta_1} \!\!\!\!\D\theta_2 \left[
 	\mean{\h{O}(x)\h{T}_{\lambda\sigma}(\theta_1,\vec{x}_1) \h{T}_{\gamma\delta}(\theta_2,\vec{x}_2)}_\beta
	-\mean{\h{T}_{\lambda\sigma}(\theta_1,\vec{x}_1) \h{T}_{\gamma\delta}(\theta_2,\vec{x}_2)\h{O}(x)}_\beta\right]\\
&+\frac{1}{2\beta^2}\int_{-\infty}^{t_1} \D\theta_1 \int_{-\infty}^{t_2}\D\theta_2\left[
	\mean{\h{T}_{\gamma\delta}(\theta_2,\vec{x}_2)\h{O}(x)\h{T}_{\lambda\sigma}(\theta_1,\vec{x}_1)}_\beta
	-\mean{\h{T}_{\lambda\sigma}(\theta_1,\vec{x}_1) \h{T}_{\gamma\delta}(\theta_2,\vec{x}_2)\h{O}(x)}_\beta\right].
\end{split}
\end{equation*}
The integral region in the second line can be split in two parts in order to recreate the region of the integral
of the first line; we find:
\begin{equation*}
\begin{split}
A_1^{C} = & -\frac{1}{2\beta^2}\!\! \int_{-\infty}^{t_1} \!\D\theta_1\!\int_{-\infty}^{\theta_1} \!\!\!\!\D\theta_2 \left[
 	\mean{\h{O}(x)\h{T}_{\lambda\sigma}(\theta_1,\vec{x}_1) \h{T}_{\gamma\delta}(\theta_2,\vec{x}_2)}_\beta
 	-\mean{\h{T}_{\gamma\delta}(\theta_2,\vec{x}_2)\h{O}(x)\h{T}_{\lambda\sigma}(\theta_1,\vec{x}_1)}_\beta	\right]\\
&-\frac{1}{2\beta^2}\!\! \int_{-\infty}^{t_2} \!\D\theta_2\!\int_{-\infty}^{\theta_2} \!\!\!\!\D\theta_1\left[
	\mean{\h{T}_{\lambda\sigma}(\theta_1,\vec{x}_1) \h{T}_{\gamma\delta}(\theta_2,\vec{x}_2)\h{O}(x)}_\beta
	-\mean{\h{T}_{\gamma\delta}(\theta_2,\vec{x}_2)\h{O}(x)\h{T}_{\lambda\sigma}(\theta_1,\vec{x}_1)}_\beta	\right].
\end{split}
\end{equation*}
Similarly, for $A_2$ we find
\begin{equation*}
\begin{split}
A_2^{C} = & -\frac{1}{2\beta^2}\!\! \int_{-\infty}^{t_2} \!\D\theta_2\!\int_{-\infty}^{\theta_2} \!\!\!\!\D\theta_1 \left[
 	\mean{\h{O}(x) \h{T}_{\gamma\delta}(\theta_2,\vec{x}_2) \h{T}_{\lambda\sigma}(\theta_1,\vec{x}_1)}_\beta
 	-\mean{\h{T}_{\lambda\sigma}(\theta_1,\vec{x}_1) \h{O}(x) \h{T}_{\gamma\delta}(\theta_2,\vec{x}_2)}_\beta	\right]\\
&-\frac{1}{2\beta^2}\!\! \int_{-\infty}^{t_1} \!\D\theta_1\!\int_{-\infty}^{\theta_1} \!\!\!\!\D\theta_2\left[
	\mean{\h{T}_{\gamma\delta}(\theta_2,\vec{x}_2) \h{T}_{\lambda\sigma}(\theta_1,\vec{x}_1) \h{O}(x)}_\beta
	-\mean{\h{T}_{\lambda\sigma}(\theta_1,\vec{x}_1)\h{O}(x)\h{T}_{\gamma\delta}(\theta_2,\vec{x}_2)}_\beta	\right].
\end{split}
\end{equation*}
It is straightforward to realize that
\begin{equation*}
\begin{split}
A_1^C + A_2^{C} = &-\frac{1}{2\beta^2}\!\! \int_{-\infty}^{t_1} \!\D\theta_1\!\int_{-\infty}^{\theta_1} \!\!\!\!\D\theta_2
 	\mean{\left[\left[\h{O}(x),\h{T}_{\lambda\sigma}(\theta_1,\vec{x}_1)\right], \h{T}_{\gamma\delta}(\theta_2,\vec{x}_2)\right]}_\beta \\
&-\frac{1}{2\beta^2}\!\! \int_{-\infty}^{t_2} \!\D\theta_2\!\int_{-\infty}^{\theta_2} \!\!\!\!\D\theta_1
	\mean{\left[\left[\h{O}(x), \h{T}_{\gamma\delta}(\theta_2,\vec{x}_2)\right],\h{T}_{\lambda\sigma}(\theta_1,\vec{x}_1)\right]}_\beta.
\end{split}
\end{equation*}
As expected from the definition of this connected correlator, the disconnected terms cancel out,
namely:
$$
A_1^{DC}+A_2^{DC}+B=0.
$$
To conclude, we showed that Eq. (\ref{eq:DeltaOmegaEtaAplusB}) is
\begin{equation*}
\Delta_{\omega\eta}O(x)= \int_{T\Sigma}\!\D^3 x_1 \!\int\!\D^4 x_2\,\, n^\lambda (x_1-x)_\rho
	\left(A_1^C + A_2^C \right)\beta(x_2)\sigma^{\mu\nu}(x_2),
\end{equation*}
and hence, renaming $\theta_1$ to $t_1$, it is equal to Eq. (\ref{eq:DeltaOmegaEtaORetardedGreen}).

\subsection{By parts integration}
\label{sec:ByParts}
In the derivation of the Kubo formula we considered
pertubation of equilibrium configurations with four-wave vector $q$.
The anomalous current is
\begin{equation*}
\begin{split}
\Delta_{\omega\eta} j^\mu(x) =& 2 \sigma^{\mu\nu}(x) \omega_\nu(x) \text{Re}
    \int_{t_0}^t \D^4 x_2 \int_{t_0}^{t_2} \D\theta_2 \int_{t_0}^t \D^4 x_1  (x_1-x)^y
    \,\I\, G^{\rm R1}_{\h{j}^y,\h{T}^{tz},\h{T}^{xy}}\left(x;\, x_1,\, (\theta_2, \vec{x}_2)\right) \E^{-\I q\cdot(x_2-x)}.
\end{split}
\end{equation*}
In the following we denote with $q_0$ the projection of $q$ along $n$, and with $\vec{q}$ its
transverse direction. Consider then
\begin{equation*}
\int_{t_0}^t \D^4 x_2 \int_{t_0}^{t_2} \D\theta_2 F(\theta_2) \E^{-\I q_0\cdot(t_2-t)}
\end{equation*}
where
\begin{equation*}
F(\theta_2) \equiv \int_{t_0}^t \D^4 x_1  (x_1)^y
    G^{\rm R1}_{\h{j}^y,\h{T}^{tz},\h{T}^{xy}}\left(x;\, x_1,\, (\theta_2, \vec{x}_2)\right) \E^{\I \vec{q}\cdot(\vec{x}_2-\vec{x})},
\end{equation*}
we also define
\begin{equation*}
G(t_2)\equiv \int_{t_0}^{t_2} \D\theta_2 F(\theta_2).
\end{equation*}
The integral in $t_2$ can be done by parts:
\begin{equation*}
\begin{split}
\int_{t_0}^t \D t_2 \, G(t_2) \E^{-\I q_0\cdot(t_2-t)} = &
    \int_{t_0}^t \D t_2 \, G(t_2) \frac{\I}{q_0}\frac{\D}{\D t_2} \E^{-\I q_0\cdot(t_2-t)}\\
= & \frac{\I}{q_0} \int_{t_0}^t \D t_2 \frac{\D}{\D t_2} \left(G(t_2)\E^{-\I q_0(t_2-t)} \right)
    - \frac{\I}{q_0} \int_{t_0}^t \D t_2 \frac{\D G(t_2)}{\D t_2} \E^{-\I q_0(t_2-t)} \\
= & \frac{\I}{q_0}\left[G(t)-G(t_0) \E^{-\I q_0 (t_0-t)}
    -\int_{t_0}^t \D t_2 \, F(t_2) \E^{-\I q_0(t_2-t)} \right]
\end{split}
\end{equation*}
If, as it is reasonable, the quantity $F(t_0)$ is not divergent, i.e. $F(t_0)<\infty$, we have
\begin{equation}
G(t_0) = \int_{t_0}^{t_0} \D\theta_2\,F(\theta_2)=0,
\end{equation}
hence
\begin{equation*}
\begin{split}
\int_{t_0}^t \D t_2 \, G(t_2) \E^{-\I q_0\cdot(t_2-t)}
= & \frac{\I}{q_0} \int_{t_0}^t \D t_2 \, F(t_2) \left( 1- \E^{-\I q_0(t_2-t)}\right)
= -\I \frac{\D}{\D q_0} \int_{t_0}^t \D t_2\, F(t_2) \left.\E^{-\I q_0 (t_2-t)}\right|_{q_0=0}.
\end{split}
\end{equation*}
We finally obtain
\begin{equation*}
\begin{split}
\Delta_{\omega\eta} j^\mu(x) =& 2 \sigma^{\mu\nu}(x) \omega_\nu(x)
    \left.\frac{\pd}{\pd q_0}\right|_{q_0=0}\lim_{\vec{q}\to 0} \text{Re}(-\I)
    \int_{t_0}^t \D^4 x_2 \int_{t_0}^t \D^4 x_1  (x_1-x)^y
    \,\I\, G^{\rm R1}_{\h{j}^y,\h{T}^{tz},\h{T}^{xy}}(x;\, x_1,\, x_2) \E^{-\I q\cdot(x_2-x)}
\end{split}
\end{equation*}
or restoring the covariant form:
\begin{equation*}
\begin{split}
\Delta_{\omega\eta} j^\mu(x) =& - 4 \beta^2 \sigma^{\mu\nu} \omega_\nu
    n^\alpha\left.\frac{\pd}{\pd q^\alpha}\right|_{q\cdot n=0}\lim_{q_T\to 0} \text{Im}
    \int_{t_0}^t \D^4 x_2 \int_{t_0}^t \D^4 x_1  (x_1-x)^y
    \,\I\, G^{\rm R1}_{\h{j}^y,\h{T}^{tz},\h{T}^{xy}}(x;\, x_1,\, x_2) \E^{-\I q\cdot(x_2-x)}.
\end{split}
\end{equation*}

\end{widetext}

\end{document}